\documentclass{aastex631}

\shorttitle{On the Sweet-Parker model}
\shortauthors{Baty}
\graphicspath{{./}{figures/}}

\usepackage{bm}%
\usepackage{graphicx}%
\usepackage{amsmath}
\usepackage{amsmath,esint}

\def\ltsima{$\; \buildrel < \over \sim \;$}
\def\gtsima{$\; \buildrel > \over \sim \;$}
\def\simlt{\lower.5ex\hbox{\ltsima}}
\def\simgt{\lower.5ex\hbox{\gtsima}}

\begin{document}
\title{
On the Sweet-Parker model for incompressible visco-resistive magnetic reconnection
in two dimensions associated to ideal magnetohydrodynamic instabilities}

\author[0000-0003-1925-3983]{Hubert Baty}
\affiliation{Observatoire Astronomique de Strasbourg,
Universit\'e de Strasbourg \\
11 Rue de l'universit\'e, 67000 Strasbourg, FRANCE \\
hubert.baty@unistra.fr}



\begin{abstract}

We revisit the well known Sweet-Parker (SP) model for magnetic reconnection in the framework of two dimensional
incompressible magnetohydrodynamics. The steady-state solution is re-derived by considering a non zero viscosity via
the magnetic Prandtl number $P_m$. Moreover, contrary to the original SP model, a particular attention is paid to the possibility that the inflowing
magnetic field $B_e$ and the length of the current layer $L$ are not necessarily fixed and may depend on the dissipation parameters. 
Using two different ideally unstable setups to form the current sheet, namely the tilt and coalescence modes, we numerically explore the scaling
relations with resistivity $\eta$ and Prandtl number $P_m$ during the magnetic reconnection phase, and compare to the generalized steady-state SP theoretical solution.
The usual Sweet-Parker relations are recovered in the limit of small $P_m$ and $\eta$ values, with in particular the normalized reconnection rate being simply $S^{-1/2} (1 + P_m)^{-1/4}$,
where $S$ represents the Lundquist number $S = LV_A/\eta$ ($V_A$ being the characteristic Alfv\'en speed).
In the opposite limit of higher $P_m$ and/or $\eta$ values, a significant deviation from the SP model is obtained with a complex
dependence $B_e (\eta, P_m)$ that is explored depending on the setup considered. We discuss the importance of these results in order to correctly interpret the numerous
exponentially increasing numerical studies published in the literature, with the aim of explaining eruptive phenomena observed in the solar corona.

\end{abstract}



\section{Introduction} \label{sec:intro}

Since its introduction, the Sweet-Parker (SP) model is considered to be the solution of reference for magnetic reconnection solution in
two-dimensional (2D) Magnetohydrodynamic (MHD) framework \citep{swe58, par57, pri00}. It assumes incompressibility of the flow and the viscosity effect is
neglected. The model also focus on a steady-state solution in presence of a pre-formed current sheet of fixed half-length $L$ and
fixed inflowing magnetic field $B_e$ far from the layer.
The results are very enlightening as the outflow velocity is simply given by the Alfv\'en speed $V_A$, based on $B_e$ with $V_A = B_e/(\rho \mu_0)^{1/2}$
(where $ \rho$ is the constant and uniform mass density, and $\mu_0$ is the vacuum magnetic permeability parameter). The half-width of the current sheet
is also given by $\delta = L S^{-1/2}$, where $S = L V_A / \eta$ is the Lundquist number ($\eta$ being the magnetic diffusivity or the resistivity parameter). Finally the
reconnection rate which measures the speed of the process (in dimensionless units using the Alfv\'en velocity for normalization) is simply given by $S^{-1/2}$.

The above scaling relations often serve as a reference in order to test MHD codes and compare with results obtained in numerical experiments.
However, all MHD codes always contain a non zero viscosity, at least due to the numerical scheme. Consequently, the original SP model 
must be modified to take into account viscosity effect, in order to allow for more precise theoretical scaling laws and consequently a correct interpretation of the numerical results.
As this is rarely considered in the literature, to the exception of the study of \citet{par84} in the context of tokamak plasmas, we propose to revisit this point in the present work.
Moreover, we relax the assumption of the independence of $L$  and $B_e$ with the dissipation parameters. For example, the dependence of $B_e$ with the
resistivity $\eta$ was shown to be important in order to interpret numerical experiments using coalescence instabilities \citep{bisk80, delu92}.

In order to test our scaling laws, we use two different setups based on an initial ideal MHD instability in order to form the current sheet.
In this way, contrary to the use of a resistive instability (as for example the tearing mode), the initial linear phase is not (or weakly)
influenced by the dissipation parameters. We choose the tilt and coalescence configurations to do so in the numerical experiments. 
We use a strongly adaptive finite-element code, FINMHD, which has been specifically designed to address such reconnection problem within the framework
of reduced visco-resistive MHD in a two-dimensional Cartesian geometry \cite{bat19}. Note that we focus on the regime where the Lundquist number $S$
is limited to values lower than a critical value $S_c$ (that is $S_c \simeq 10^4$ in the limit of vanishing viscosity), in order to exclude 
the stochastic reconnection regime dominated by the presence of plasmoids \citep{lou07, com16, bat20a, bat20b, bat20c}.

The outline of the paper is as follows. In Section 2, we derived the reconnection solution with generalized SP scaling laws including the
viscosity effect.  We briefly present the code and the initial setups in Section 3. Section 4 is devoted to the presentation of the results. Finally, we conclude in Section 5.

\section{Generalized Sweet-Parker (SP) reconnection model}

\begin{figure}
\centering
 \includegraphics[scale=0.32]{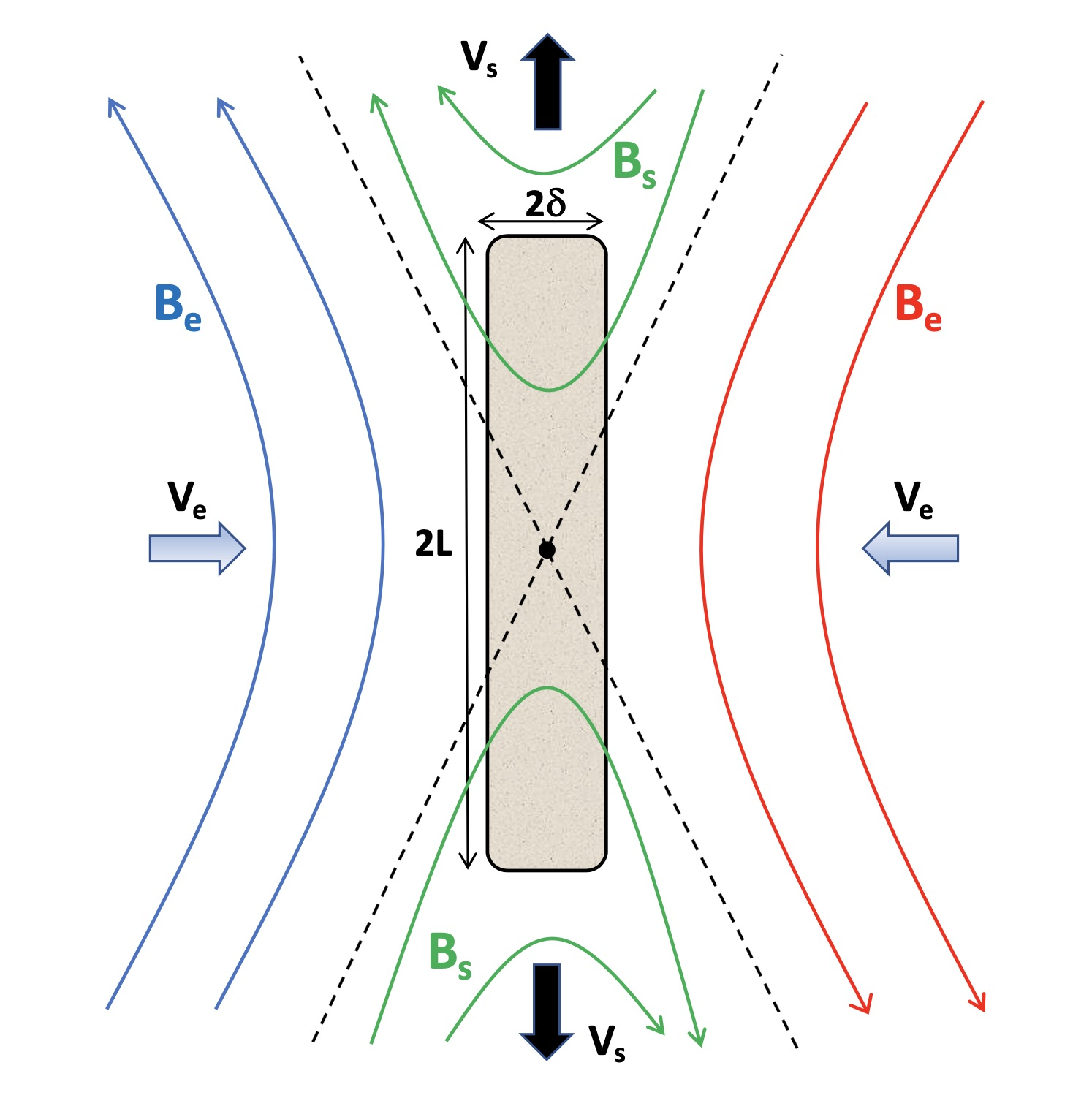}
  \caption{Schematic view of the standard Sweet-Parker reconnection model in a two-dimensional $(x-y)$-plane.
 Magnetic field lines of magnitude ($B_e$ in blue and red) are advected at an inflowing speed $V_e$ (along the $x$-axis) towards a central diffusive region of dimensions $2 \delta \times 2 L$ (defining the current sheet layer) where the direction of the magnetic field has a reversal. The reconnected field lines of magnitude ($B_s$ in green) are expelled
  from the centre (e.g. $X$-point) and accelerated along the current sheet (i.e. $y$-axis), reaching an outflow speed $V_s$.
  }
\label{fig1}
\end{figure}

The schematic structure of the SP configuration is visible in Figure 1. We consider a current sheet having a full length $2L$ and a full thickness $2\delta$.
The first equation used is the incompressibility condition of the flow $\nabla \cdot \bm{V} = 0$, leading thus to $ \oiint\limits_{S} \bm{V} \cdot \bm{dS} = 0$ where
the integral is taken over the surface $S$ enclosing the current sheet volume. Consequently, the first relation that is in fact the mass flux conservation between entrance and exit of the current sheet is,
\begin{equation}  
      V_e L = V_s \delta ,
\end{equation}
with $V_s$ denoting the outflow velocity of the reconnected field lines ($V_e$ being the inflow speed).

The second equation used can be deduced from Faraday's law $ \bm{\nabla} \times  \bm{E} =  - \frac{\partial B}{\partial t}$, and Ohms's law $ \bm{E} = - \bm{V} \times \bm{B} + \mu_0 \eta  \bm{J} $
for the electric field $ \bm{E}$. Assuming steady-state thus leads to a uniform value for the electric field component perpendicular to the plane $E_z = - (\bm{V} \times \bm{B})_z + \mu_0 \eta J_z$.
The value of $E_z$ in the inflowing region must be thus equal to the value in the outflowing region, that are moreover considered to be regions where the resistive term ($\eta J_z$) is negligible,
leading to the second relation,
\begin{equation}  
     E_z = V_e B_e = V_s B_s,
\end{equation}
that is in fact a magnetic flux conservation between entrance and exit of the current layer. Note that the resistivity parameter $\eta$ used here, denotes the magnetic diffusivity ($m^2/s$ in MKSA units)
and does not encompass $\mu_0$ as sometimes chosen in other studies.

Using the dominance of the resistive term over the ideal one inside the current layer (as the magnetic field is reversing at the center), one gets another estimate for the electric field that is $E_z = \mu_0 \eta J_z$ which can be approximated by,
\begin{equation}  
     E_z \simeq \eta B_e / \delta ,
   \end{equation}
using Ampere's law $  \bm{\nabla} \times  \bm{B}  =  \mu_0 \bm{J} $ and assuming a very small thickness $\delta$ compared to $L$ (to be a posteriori checked). A symmetric reversal is also taken
for the sake of simplification.

The dynamics of the process is determined by the momentum conservation equation, which can be written in a steady-state form (neglecting the thermal pressure gradient),
$(\bm{V}\cdot\bm{\nabla}) \bm{V}  =  \frac {\bm{J}  \times  \bm{B} }  {\rho}  + \nu {\nabla}^2 \bm{V} $, where $\rho$ denotes the uniform mass density. As done in the original
SP model, one can integrate this equation along the current sheet between the centre ($y = 0$) and the exit ($y = L$) assuming a linear variation of the outflowing fields
(i.e. $V_y$ and $B_x$ components). The centre of the 2D cartesian frame is taken at the $X$-point.
Equivalently, one can evaluate the different terms at mid-distance ($y = L/2$) by taking the average values $V_s/2$ and $B_s/2$ for the velocity and magnetic field respectively.
The current density term in the magnetic force is however nearly constant along the current sheet (checked in simulations) and is consequently
evaluated by its maximum value $J_M$ taken at the centre $\mu_0 J_M  \simeq B_e / \delta$. Consequently, we obtain
$  \frac{V_s} {2}   \frac{(V_s - 0)}  {L} \simeq \frac{J_M B_s} {2 \rho}  - \nu  \frac{V_s} {2 \delta^2} $, leading to the fourth relation,
\begin{equation}  
      V_s (V_s + \nu  \frac{L} {\delta^2})  \simeq \frac{ B_s B_e} {\rho \mu_0 }  \frac{L} {\delta} .
   \end{equation}

Combining Eq.1 and Eq.2, one can easily check that $B_e = B_s L/\delta$. Moreover, using Eqs. 1-3, we have $L/\delta^2 \simeq V_s/ \eta$.
Inserting these two results in the fourth above relation, one can get the important expression for the outflow velocity,
\begin{equation}  
       V_s \simeq \frac{B_e} {(\rho \mu_0 )^ {1/2} } \frac{1}  {(1 + \nu/\eta)^ {1/2} } .
   \end{equation}
Note that the Alfv\'en speed $V_s \simeq V_A = \frac{ B_e} {(\rho \mu_0)^{1/2}}$
of the standard inviscid SP model is obviously recovered in the limit of zero viscosity (i.e. vanishing magnetic Prandtl number), and that the outflow is slowed down by a
factor $(1 + P_m)^ {1/2}$ by the viscous force (as $P_m$ is defined as $P_m = \nu/\eta$). The latter result was previously derived by \citet{par84} in the context of magnetic reconnection in tokamak plasmas.

As a consequence, the important results (useful for the present work) can be derived as,
\begin{equation} 
\left\{
    \begin{aligned}
  &  V_s \simeq V_A (1 + P_m)^{-1/2} =   \rho^ {-1/2} \mu_0^ {-1/2}  B_e (1 + P_m)^{-1/2} \\
      &    \delta  \simeq   \rho^ {1/4} \mu_0^ {1/4} L^ {1/2}  B_e^{-1/2}  \eta^ {1/2}  (1 + P_m)^{1/4}                  \\
       &    J_M \simeq   \rho^ {-1/4} \mu_0^ {-5/4} L^ {-1/2}  B_e^{3/2}  \eta^ {-1/2}  (1 + P_m)^{-1/4}        \\
           &  \Omega_M   \simeq   \rho^ {-3/4} \mu_0^ {-3/4} L^ {-1/2}  B_e^{3/2}  \eta^ {-1/2}  (1 + P_m)^{-3/4}        ,
         \end{aligned}
  \right.
\end{equation}
where the maximum associated vorticity $\Omega_M$ is estimated via $\Omega_M \simeq V_0/\delta$. One may note that, the expected scalings using the Lundquist number $S$, $\delta /L \propto S^{-1/2}  (1+P_m)^{1/4}$, $J_M \propto S^{1/2}  (1+P_m)^{-1/4}$, and $\Omega_M  \propto S^{1/2}  (1+P_m)^{-3/4}$ can be also deduced. The reconnection rate is also an important parameter, that can be defined by using
the electric field $E_z = \mu_0 \eta J_M$, or its normalized value $E_z/(V_A B_e)$ (that is also equivalently given by the inflow Mach number $V_e/V_A$), leading thus to,
\begin{equation} 
\left\{
    \begin{aligned}
      &  E_z   \simeq   \rho^ {-1/4} \mu_0^ {-1/4} L^ {-1/2}  B_e^{3/2}  \eta^ {1/2}  (1 + P_m)^{-1/4}   ,     \\
      &     E_z/(V_A B_e)   \simeq   \rho^ {1/4} \mu_0^ {1/4} L^ {-1/2}  B_e^{-1/2}  \eta^ {1/2}  (1 + P_m)^{-1/4}  .              
               \end{aligned}
  \right.
\end{equation}
We have deliberately expressed the results in the above expressions in function of $L$ and $B_e$, which are not necessarily taken to be constant in this study.
This is not the case of the parameters $\rho$ and $\mu_0$ which can be taken to be constant and equal to unity in the following.

\section{FINMHD code and initial setups}

\begin{figure}
\centering
 \includegraphics[scale=0.18]{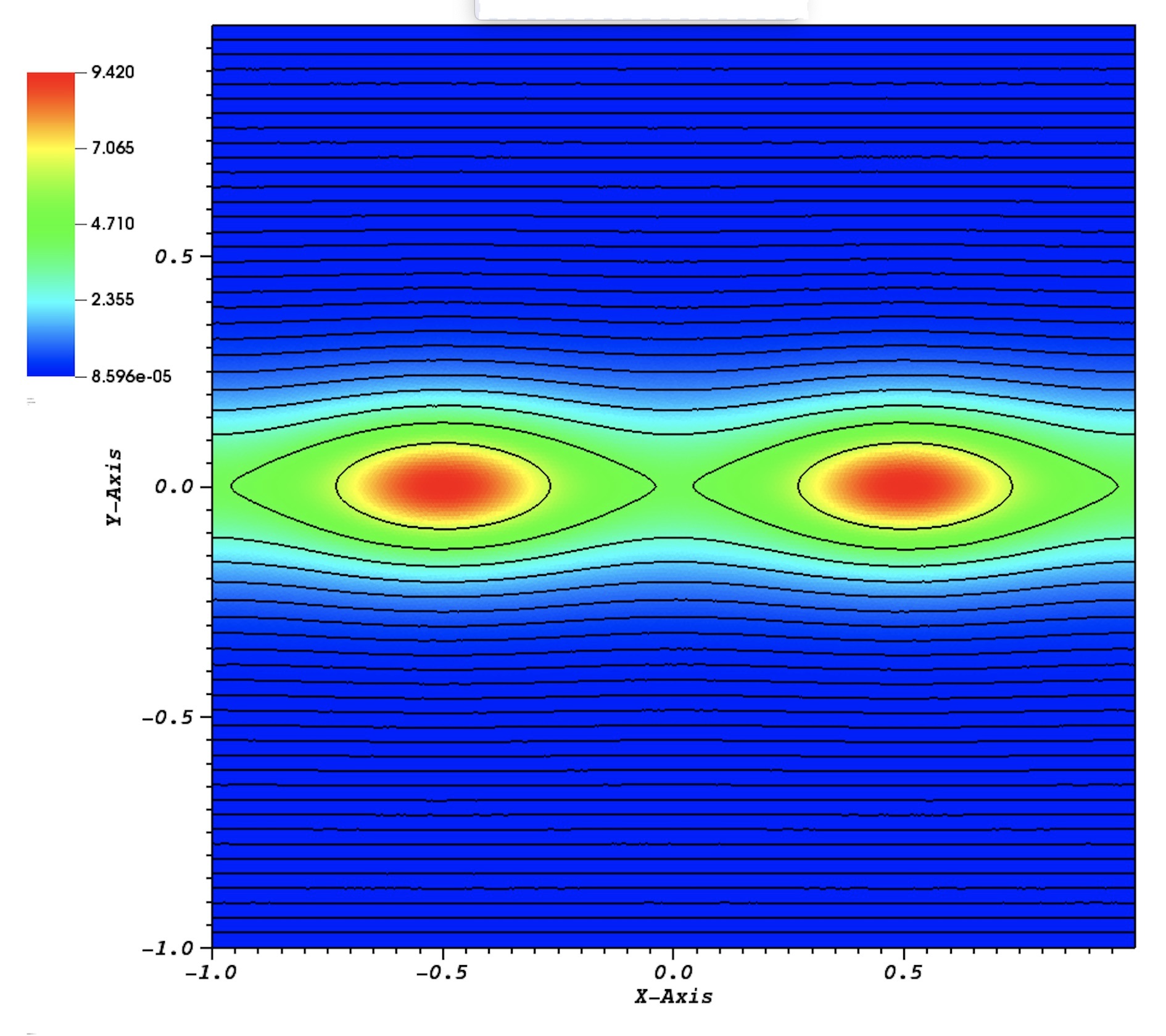}
 \includegraphics[scale=0.181]{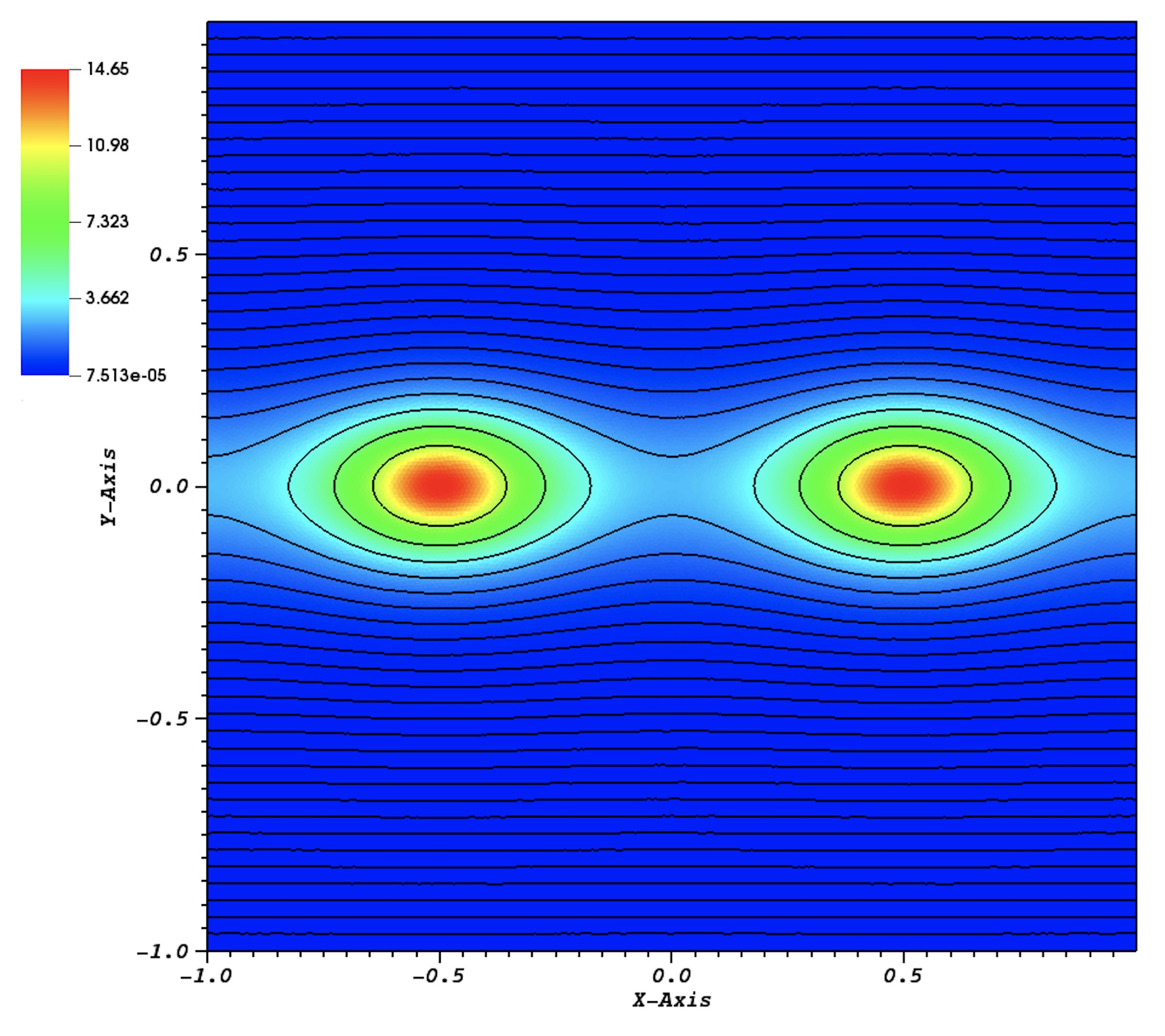}
  \caption{Initial configuration of the Fadeev equilibrium (or coalescence setup) for the current density (with colored iso-contour values) overlaid with associated magnetic field lines for two values of the $\epsilon$ parameter i.e. $\epsilon = 0.2$ and $\epsilon = 0.4$ for the left and right panel respectively. The other chosen parameters are $B_0 = 1$, $k = 2 \pi$, and $\alpha = 0$.
  }
\label{fig2}
\end{figure}
    
\begin{figure}
\centering
 \includegraphics[scale=0.25]{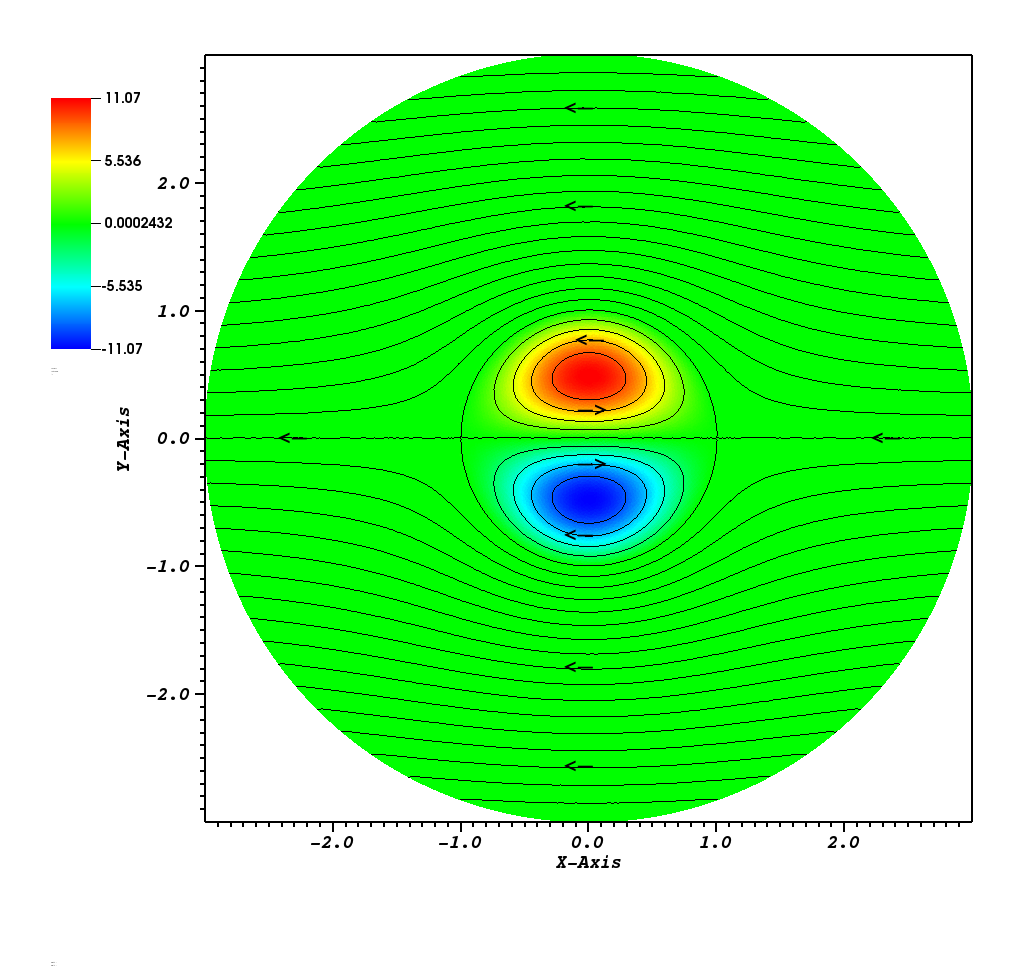}
  \caption{Initial configuration of the magnetic dipole equilibrium (or tilt setup) for the current density (with colored iso-contour values) overlaid with associated magnetic field lines.
  The chosen parameters are $B_0 = 1$ and $R = 1$, and the configuration has an external region extending to the outer boundary situated at $ r = 3$.}
\label{fig3}
\end{figure}

\subsection{FINMHD code}

The usual set of reduced MHD equations in two dimensions (2D) (i.e. $x-y$ plane) is generally admitted to be a good approximation to represent the dynamics in a plane perpendicular to a dominant 
and constant magnetic field component ($B_z$). As a consequence, the incompressibility assumption in the 2D plane is considered to be well justified. 
In this work, we use the reduced MHD formulation with two scalar variables like stream functions (hereafter 
$\phi$ and $\psi$), as this automatically ensures the divergence-free property for the corresponding plasma velocity and magnetic field vectors ($\bm V$ and $\bm B$ respectively).
Moreover, in order to facilitate the numerical implementation,
a dimensionless model using the electric current density $J$ and the flow vorticity $\Omega$ for the main variables is adopted in FINMHD \citep{bat19},
\newpage
\begin{equation}  
      \frac{\partial \Omega}{\partial t} + (\bm{V}\cdot\bm{\nabla})\Omega = (\bm{B}\cdot\bm{\nabla})J + \nu \bm{\nabla}^2 \Omega ,
\end{equation}
\begin{equation}      
        \frac{\partial J }{\partial t} + (\bm{V}\cdot\bm{\nabla})J =  (\bm{B}\cdot\bm{\nabla})\Omega + \eta \bm{\nabla}^2 (J -J_e) +  g(\phi,\psi) ,
\end{equation}
\begin{equation}                 
     \bm{\nabla}^2\phi = - \Omega ,
 \end{equation}
\begin{equation}                        
     \bm{\nabla}^2\psi = - J ,  
\end{equation}
with $g(\phi,\psi)=2 \left[\frac{\partial^2 \phi}{\partial x\partial y}\left(\frac{\partial^2 \psi}{\partial x^2} - \frac{\partial^2 \psi}{\partial y^2}\right) - \frac{\partial^2 \psi}{\partial x\partial y}\left(\frac{\partial^2 \phi}{\partial x^2} - \frac{\partial^2 \phi}{\partial y^2}\right)\right]$. We
have introduced the two stream functions, $\phi (x, y)$ and $\psi (x, y)$, defined as $\bm{V} = {\nabla} \phi \wedge \bm{e_z}$ and $\bm{B} = {\nabla} \psi \wedge \bm{e_z}$ ($\bm{e_z}$
being the unit vector perpendicular to the $xOy$ simulation plane).
Note that $J$ and $\Omega$ are the $z$ components of the current density and vorticity vectors, as $\bm{J} = \nabla \wedge \bm{B}$ and $\bm{ \Omega} = \nabla \wedge \bm{V}$ respectively (with units using $\mu_0 = 1$). Note also that we consider the resistive diffusion via the $\eta \bm{\nabla}^2 J $ term ($\eta$ being the resistivity assumed uniform for simplicity), and also a viscous term
$\nu \bm{\nabla}^2 \Omega$ in a similar way (with $\nu$ being the viscosity parameter).
The above definitions results from the choice $\psi \equiv A_z$, where $A_z$ is the $z$ component of the potentiel vector $\bm{A}$ (as $\bm{B} = \nabla \wedge \bm{A}$).
FINMHD code is based on a finite element method using triangles with quadratic basis functions on an unstructured
grid. A characteristic-Galerkin scheme is chosen in order to discretize in a stable way the Lagrangian derivatives
appearing in the two first equation. Moreover, a highly adaptive (in space and time) scheme
is developed in order to follow the rapid evolution of the solution, using either a first-order time integrator (linearly unconditionally stable) or a second-order one (subject to a CFL time-step restriction). Typically, a new adapted grid can be computed at each time step, by searching the grid that renders an estimated error nearly uniform. More precisely, the method allows to cover the current structures with
a few tens of triangles at any time, by using the Hessian matrix of the current density as the main refinement parameter.
The technique used in FINMHD has been tested on challenging tests, involving unsteady strongly anisotropic solution for the advection equation, formation of shock structures for viscous Burgers equation, and magnetic reconnection for the reduced set of MHD equations. 
The reader should refer to \citet{bat19} for more details on the numerical scheme and also to the following references for applications to different aspects of
magnetic reconnection in MHD framework \citep{bat20a, bat20b, bat20c}.

\subsection{The two initial setups}

In the previous section, the current sheet is assumed to be preformed. In order to get a more realistic configuration, the process of formation of the current layer must be included.
On the other hand, the non linear development of ideal MHD instabilities are known to be an efficient mechanism to form such layers in different plasmas (e. g. solar corona and
tokamaks). Two different well known 2D setups associated to the coalescence and tilt instabilities are thus considered in this study in order to test the scaling laws derived
in the previous section.

\begin{itemize}
 \item The coalescence setup
 \newline
The first setup represents a chain of neighboring magnetic islands in equilibrium. As initially demonstrated by using an energy principle calculation,
such configuration is unstable leading to the coalescence of the islands in a pairwise way \citep{finn77, pri79, bon83}.
This is a current driven mode due to the tendency to attract between
two currents of same sign flowing in the interior of two corresponding adjacent islands. Different choices of magnetic configuration have been done in the literature depending
mainly on the use of boundary conditions (non periodic versus singly periodic or doubly periodic) available in the subsequent numerical treatment. In the present study,
we consider the Fadeev equilibrium, defined by the equilibrium flux function $\psi_e$,
\begin{equation}      
      \psi_e (x, y) = -    \frac  {B_0}  { k}   \ln [\cosh (ky) + \epsilon \cos (kx + \alpha)].
      \end{equation}
      where $k$ and $\epsilon$ (with $0 < \epsilon < 1$) are real parametrization parameters, and $B_0$ being a magnetic field normalization magnitude. Note that
      we have added a non zero arbitrary phase parameter $\alpha$ for the seek of generality.
      The above expression follows from the force balance equation condition \citep{finn77},
      \begin{equation}      
        \nabla^2 \psi_e + f(\psi_e) = 0,
      \end{equation}
    with the particular choice $f(\psi_e) = B_0 k ( \epsilon^2 - 1) e^{2 k \psi_e/B_0} $ that is also the opposite value of the equilibrium current density, i.e. $- J_e$. One can check the corresponding
    equilibrium magnetic field components as,
    \begin{equation}      
      B_x (x, y) = -    \frac  {B_0 \sinh (ky)   }  {  \cosh (ky) + \epsilon \cos (kx + \alpha) } ,
      \end{equation}
 \begin{equation}      
      B_y (x, y) = -    \frac  {\epsilon  B_0 \sin (kx) }  {  \cosh (ky) + \epsilon \cos (kx + \alpha) } ,
      \end{equation}
and the equilibrium current density expression,   
      \begin{equation}      
      J_e (x, y) = B_0 (1 - \epsilon^2)   \frac  {k } { [ \cosh (ky) + \epsilon \cos (kx + \alpha) ]^2 }. 
           \end{equation}
           The configuration is illustrated in Figure 2 for two values of $\epsilon$, which is a measure of the width of the islands $w$, as more precisely it can be
           easily shown that $w k \simeq 4 \epsilon^{1/2}$. One must note that, a thermal pressure gradient is required in order to have a 2D MHD equilibrium
           in the momentum conservation for a standard MHD model with the velocity flow implementation. However, this
           is not needed in our 2D reduced MHD model with vorticity implementation.

\item The tilt setup
 \newline
The initial magnetic field configuration for tilt instability is a dipole current structure similar to the dipole vortex flow pattern in
fluid dynamics, where the vorticity is replaced by the current density \citep{ri90}.
It consists of two oppositely directed currents embedded in a background current-free magnetic field 
with uniform amplitude at infinitely large distance.
Contrary to the coalescence instability based on attracting parallel current structures, the two antiparallel currents in the configuration tend to repel.
The initial equilibrium is thus defined by taking the following magnetic flux distribution,
\begin{equation}
    \psi_e (x, y)=
    \left\{
      \begin{aligned}
       B_0 &\left(\frac{R^2}{r} - r\right)\frac{y}{r} ~~~& & if ~~ r > R , \\
        &- B_0 \frac{2}{\alpha J_0(\alpha R)}J_1(\alpha r)\frac{y}{r} ~~~& & if ~~ r\leq R .\\
      \end{aligned}
      \right.
  \end{equation}
  
 And the corresponding current density is,
       \begin{equation} 
    J_e (x, y) =
    \left\{
      \begin{aligned}
        &~~~~~~~~~~~~0 ~~~& & if ~~ r > R , \\
        &- B_0 \frac{2\alpha}{J_0(\alpha R)}J_1(\alpha r)\frac{y}{r} ~~~& & if ~~ r\leq R ,\\
      \end{aligned}
    \right.
\end{equation}

\noindent where 
$r=\sqrt{x^2+y^2}$, and $J_0$ et $J_1$  are Bessel functions of order $0$ and $1$ respectively.
Note also that $\alpha R$ is the first (non zero) root of $J_1$, i.e. $\alpha R \simeq 3.83170597$. The configuration is illustrated in Figure 3 for the chosen parameters $R = 1$
and $B_0 = 1$, with an outer boundary limit situated at $r = 3$. A circular outer boundary is chosen in this study in order to facilitate the numerical treatment, and  it is also
placed sufficiently far enough away from the dipole region in order to have a negligible effect on the dynamics.
As for coalescence setup, this dipole structure requires a thermal pressure in order to be in 2D equilibrium,
or alternatively an additional magnetic field component $B_z$ having a $(x, y)$ dependence is needed for a 2.5D equilibrium.
\medskip

\end{itemize}

In this work, we don't impose any initial specific perturbation, as we let the instabilities develop from the numerical noise introduced by the scheme.
For the Fadeev equilibrium setup, the boundary conditions are chosen to be periodic in $x$ direction and Dirichlet-like in the $y$ direction (with values imposed
on the different variables to be equal to their initial values). For the tilt equilibrium setup, similar Dirichlet-like conditions
are also imposed at the external radial boundary (see also \citet{bat19}). Convergence on our numerical procedure used with FINMHD code with
in particular the presentation of the adaptive (in time and space) method can be found elsewhere \citep{bat19, bat20a}.

\newpage

\section{Results}

       \begin{figure}
     \centering
 \includegraphics[scale=0.22]{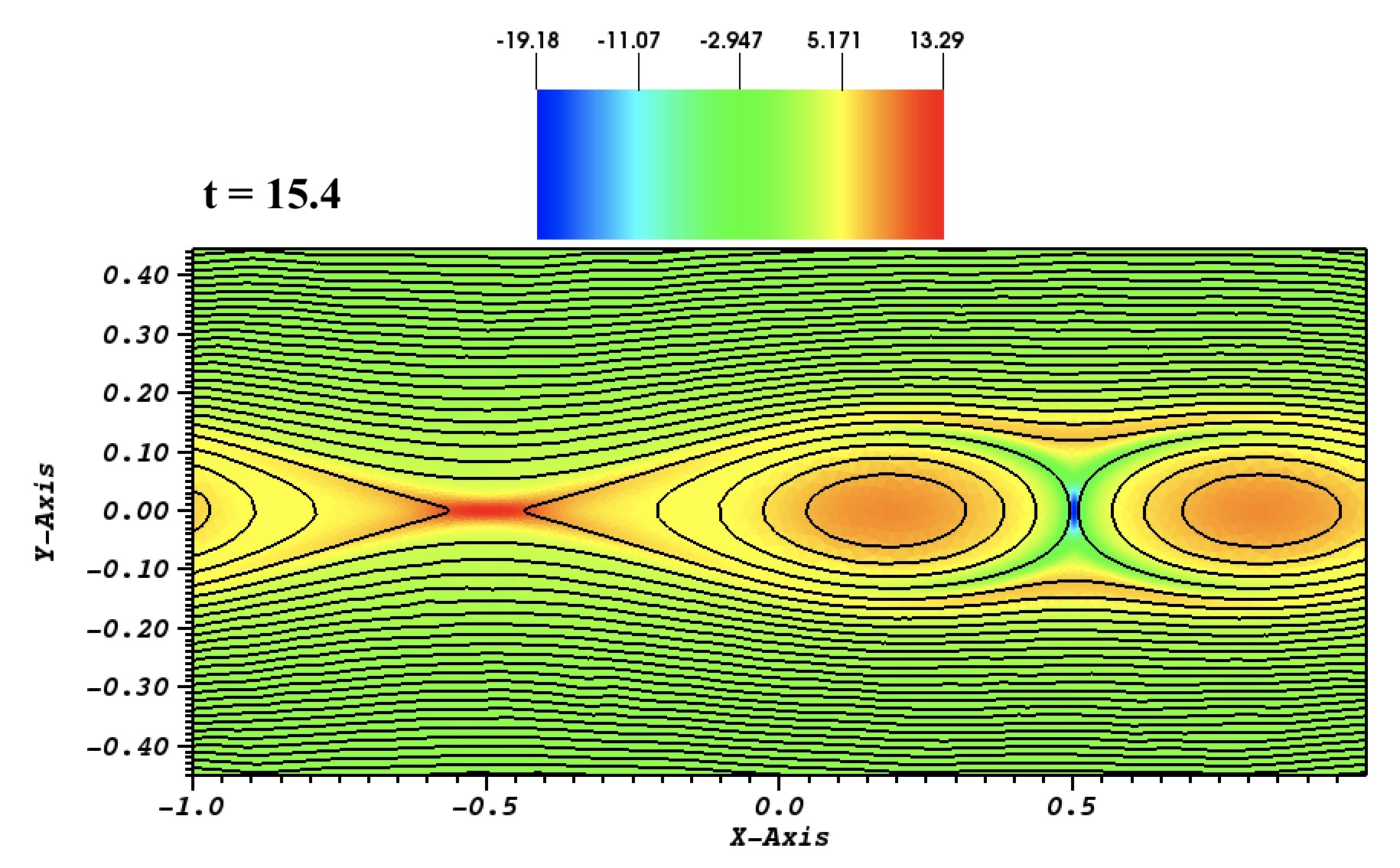}
 \includegraphics[scale=0.22]{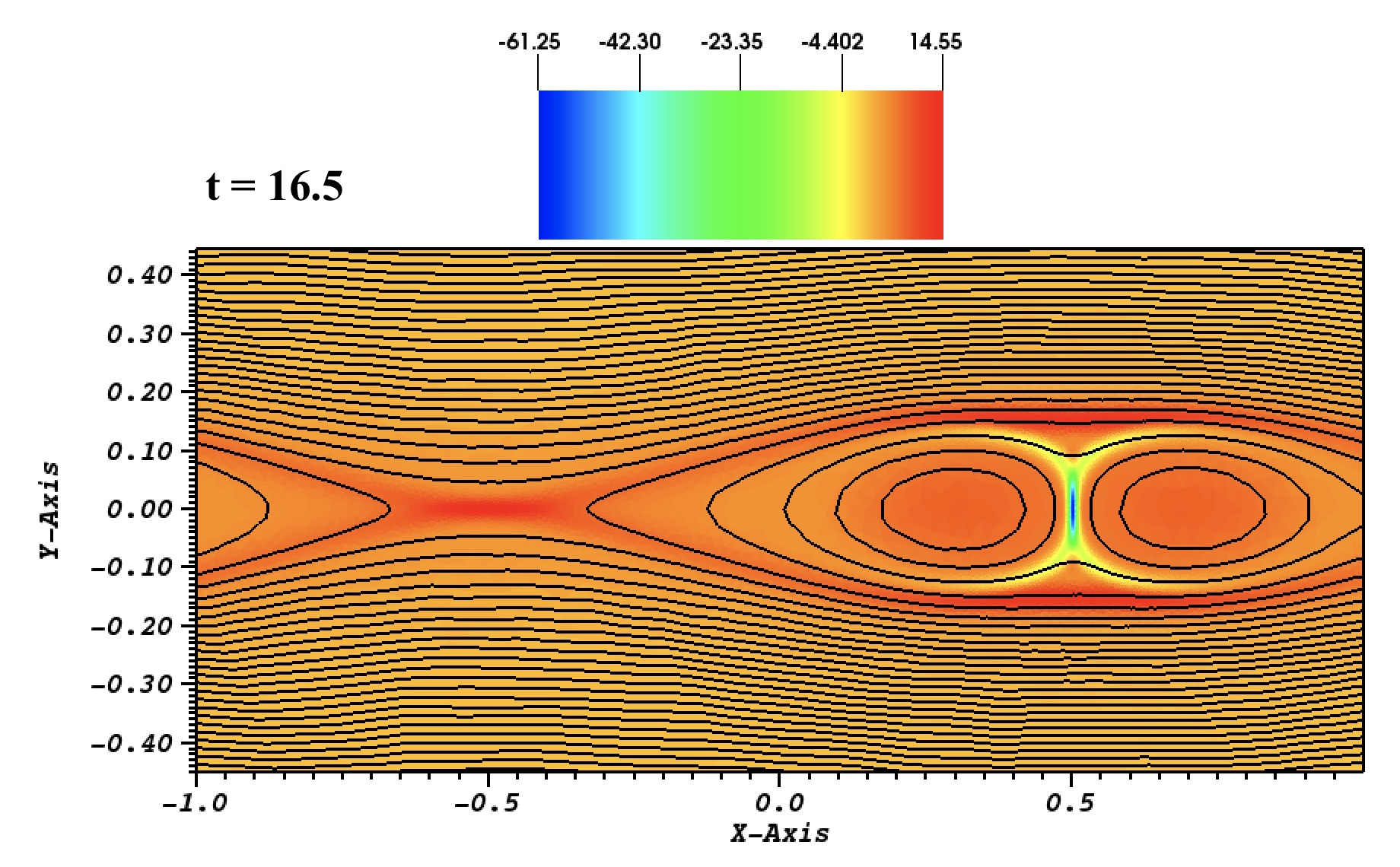}
  \includegraphics[scale=0.22]{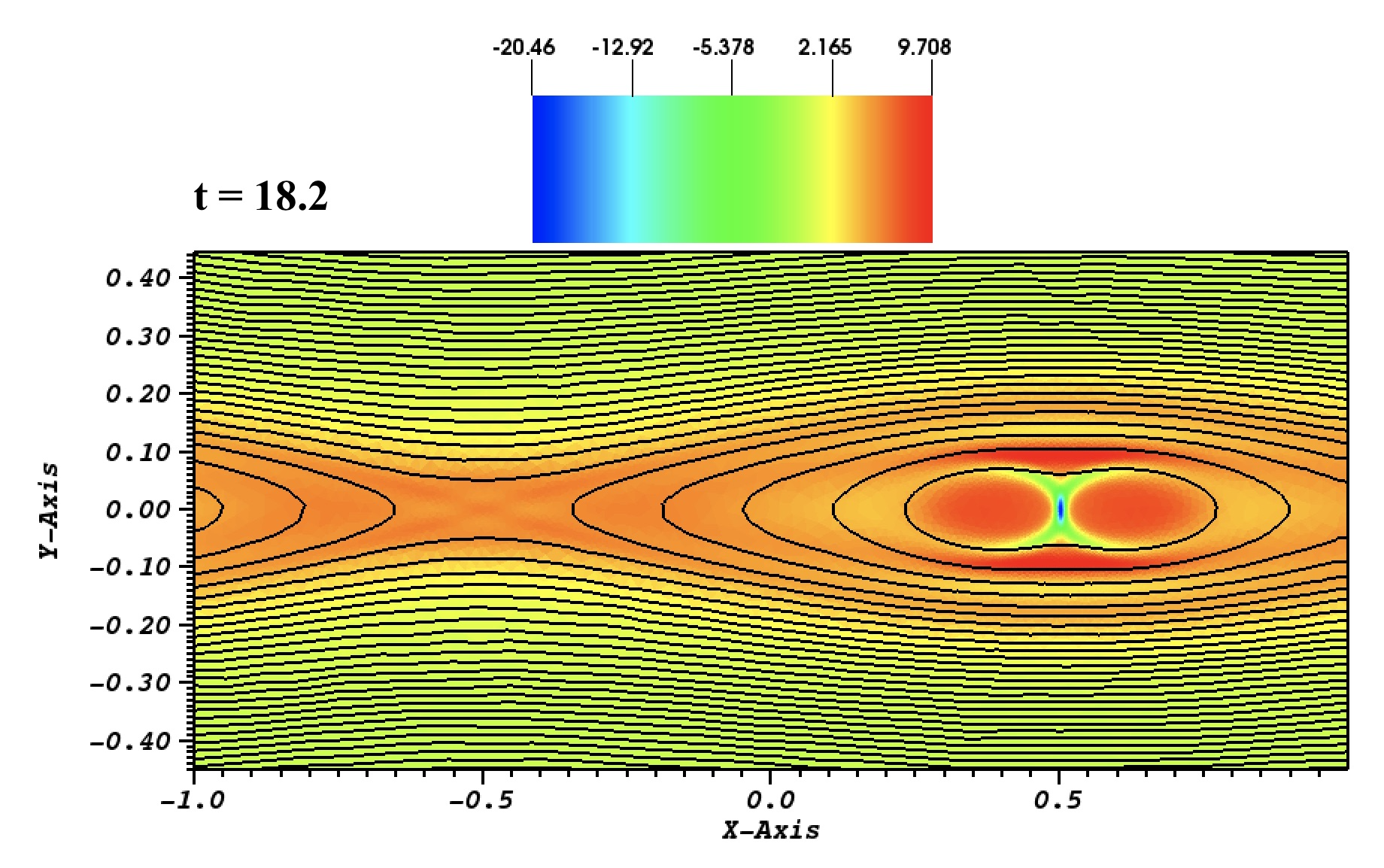}
   \includegraphics[scale=0.22]{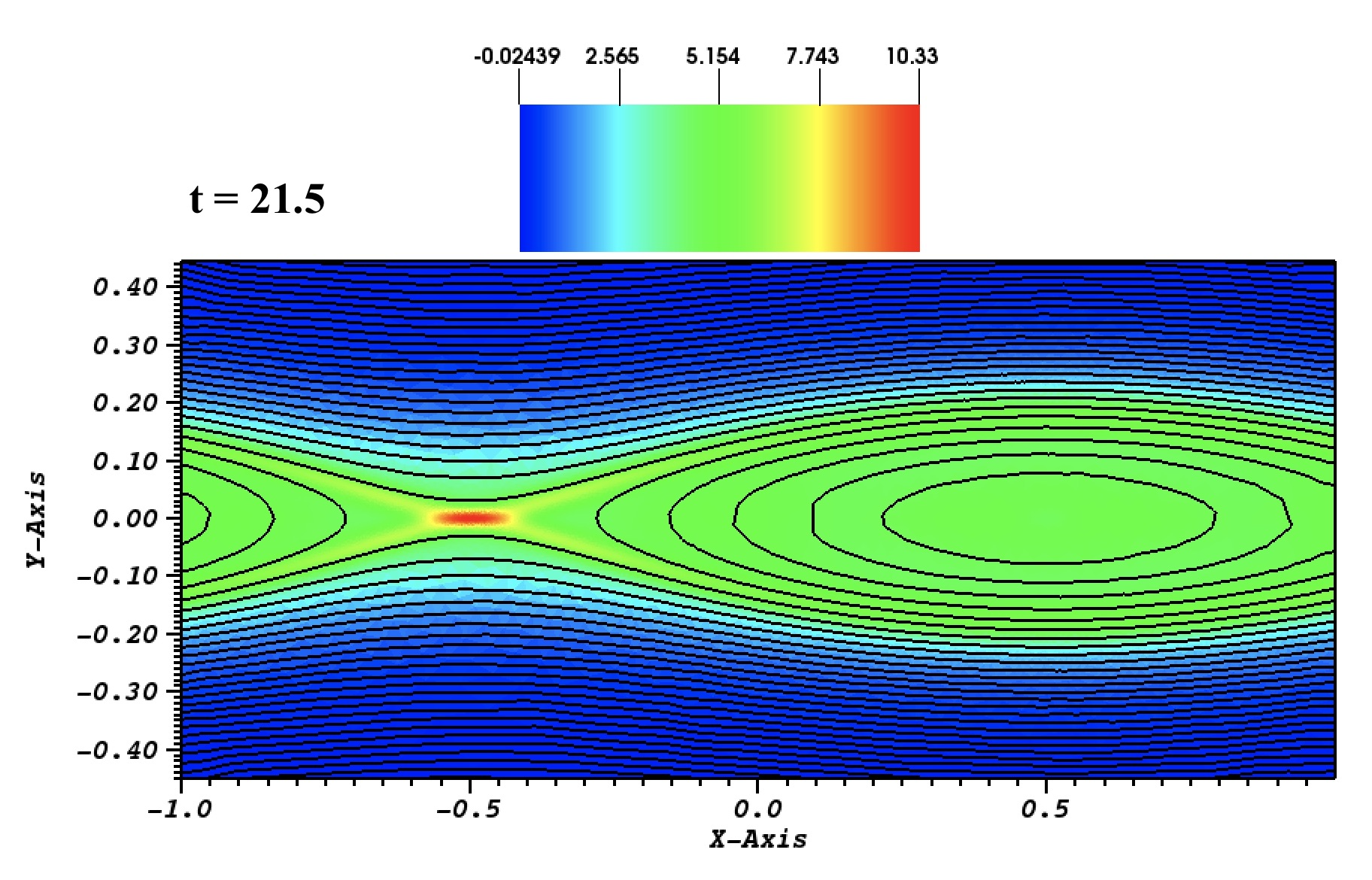}
    \caption{Snapshots taken at different times of the current density (colored iso-contours) overlaid with magnetic field lines. The run is obtained for
    the coalescence setup (or Fadeev equilibrium) using $\epsilon = 0.2$, $\alpha = \pi/2$ (leading to a current sheet localized at $x = 0.5$), and $\eta = \nu = 3.2  \times 10^{-4}$.
     }
   \label{fig4}
\end{figure}

    \begin{figure}
     \centering
 \includegraphics[scale=0.24]{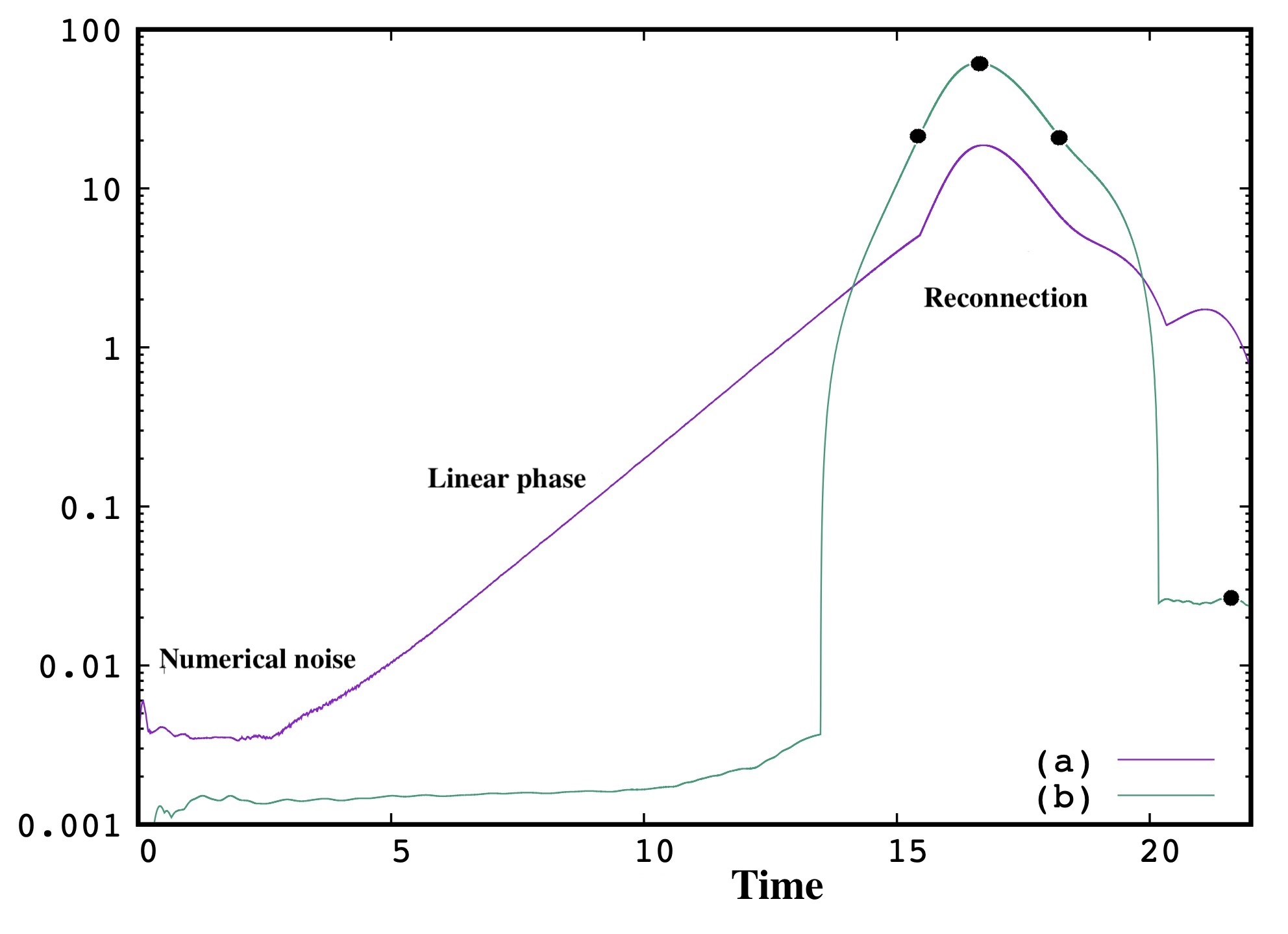}
    \caption{Maximum vorticity $\Omega_M$ (a-curve) and maximum current density measured at the $X$-point $J_M$ (b-curve) as a function of time,
    fo the run corresponding to the previous figure (Fadeev equilibrium) using $\epsilon = 0.2$, $\alpha = \pi/2$, and $\eta = \nu = 3.2  \times 10^{-4}$ (i.e. $P_m = 1$).
        }
  \label{fig5}
\end{figure}

 \begin{figure}
     \centering
 \includegraphics[scale=0.4]{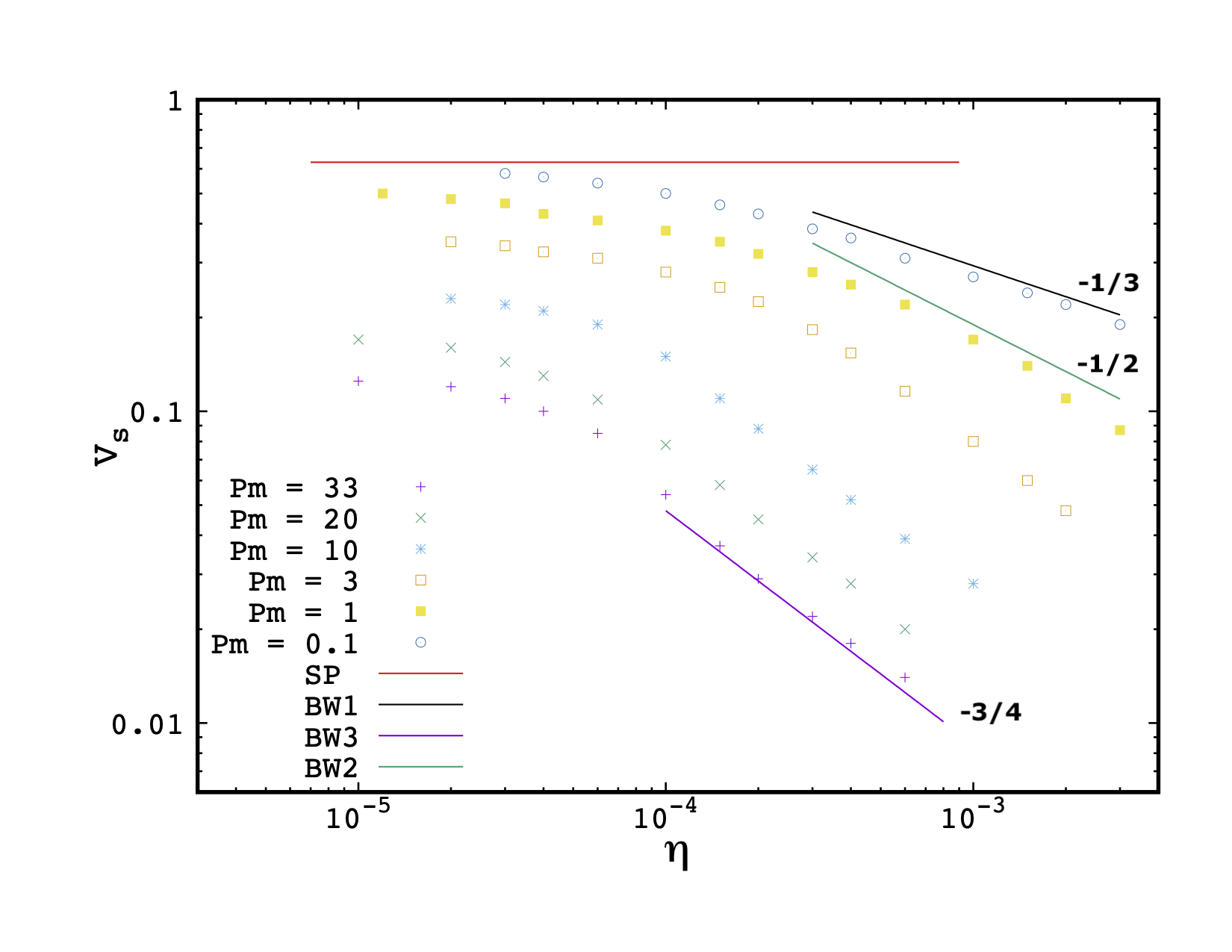}
 \includegraphics[scale=0.32]{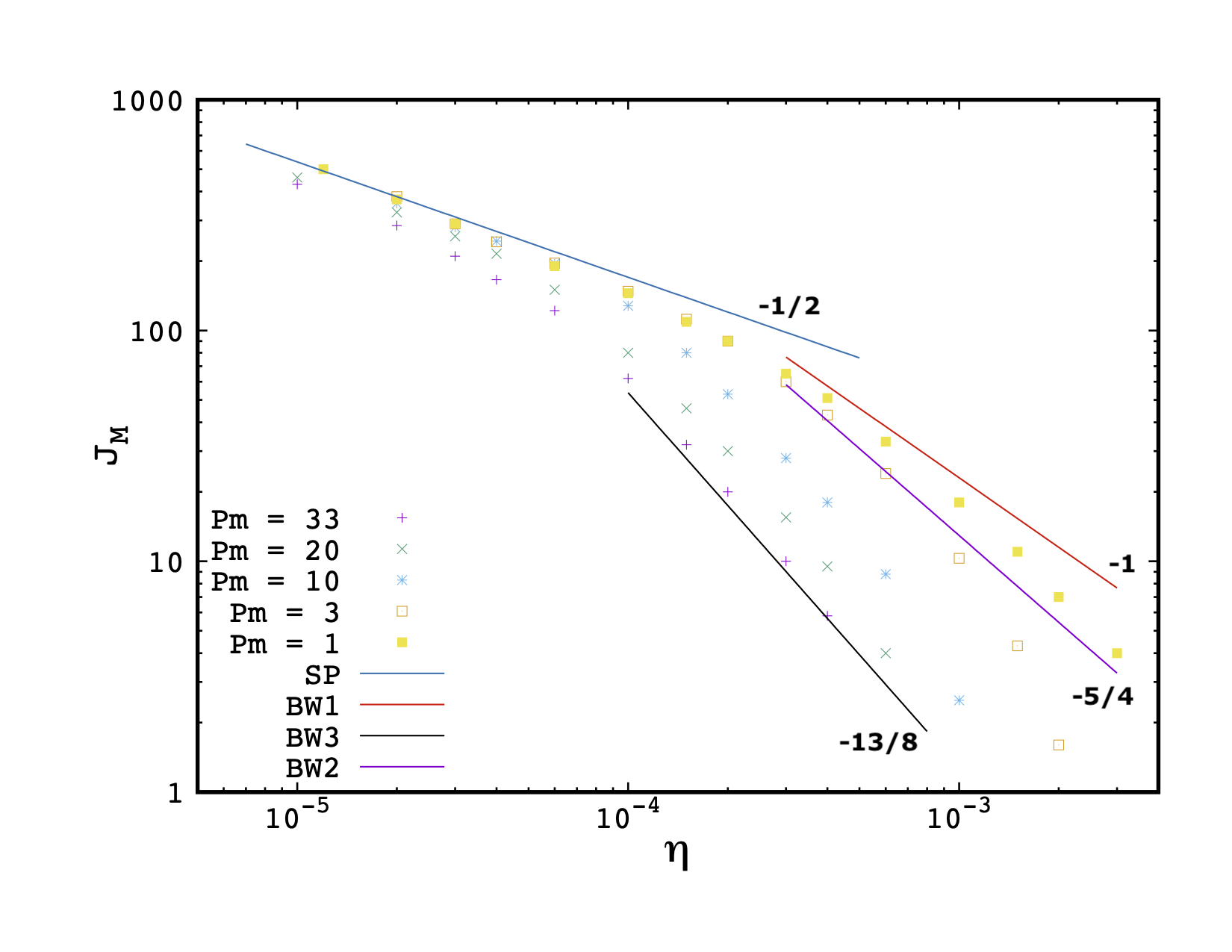}
  \includegraphics[scale=0.32]{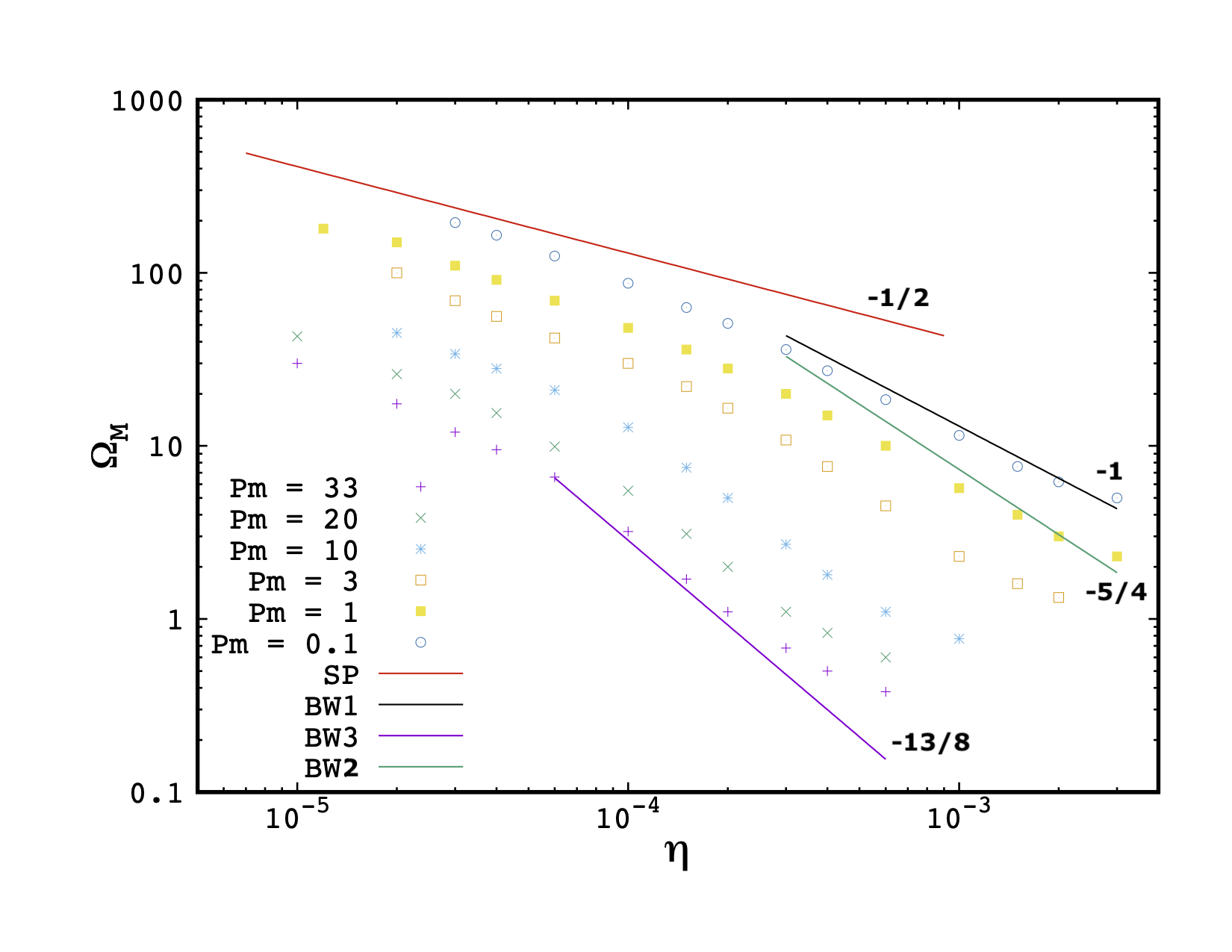}
    \caption{Scaling study of different parameters ($V_s$, $J_M$, and $\Omega_M$) with resistivity parameter $\eta$ deduced for many runs of
    the coalescence setup at different magnetic Prandtl $P_m$, i.e. 
    $P_m = 0.1, 1, 3, 10, 20$, and $33$. Sweet-Parker-like (SP) scalings, and Biskamp-Welter-like (BW) scaling laws with $\eta^{-\alpha}$ for $V_s$
    ($\alpha$ value varying in the range $[1/3 : 3/4]$) are plotted for comparison (see also text) for the BW1-2-3.
     }
   \label{fig6}
\end{figure}

 \begin{figure}
     \centering
 \includegraphics[scale=0.42]{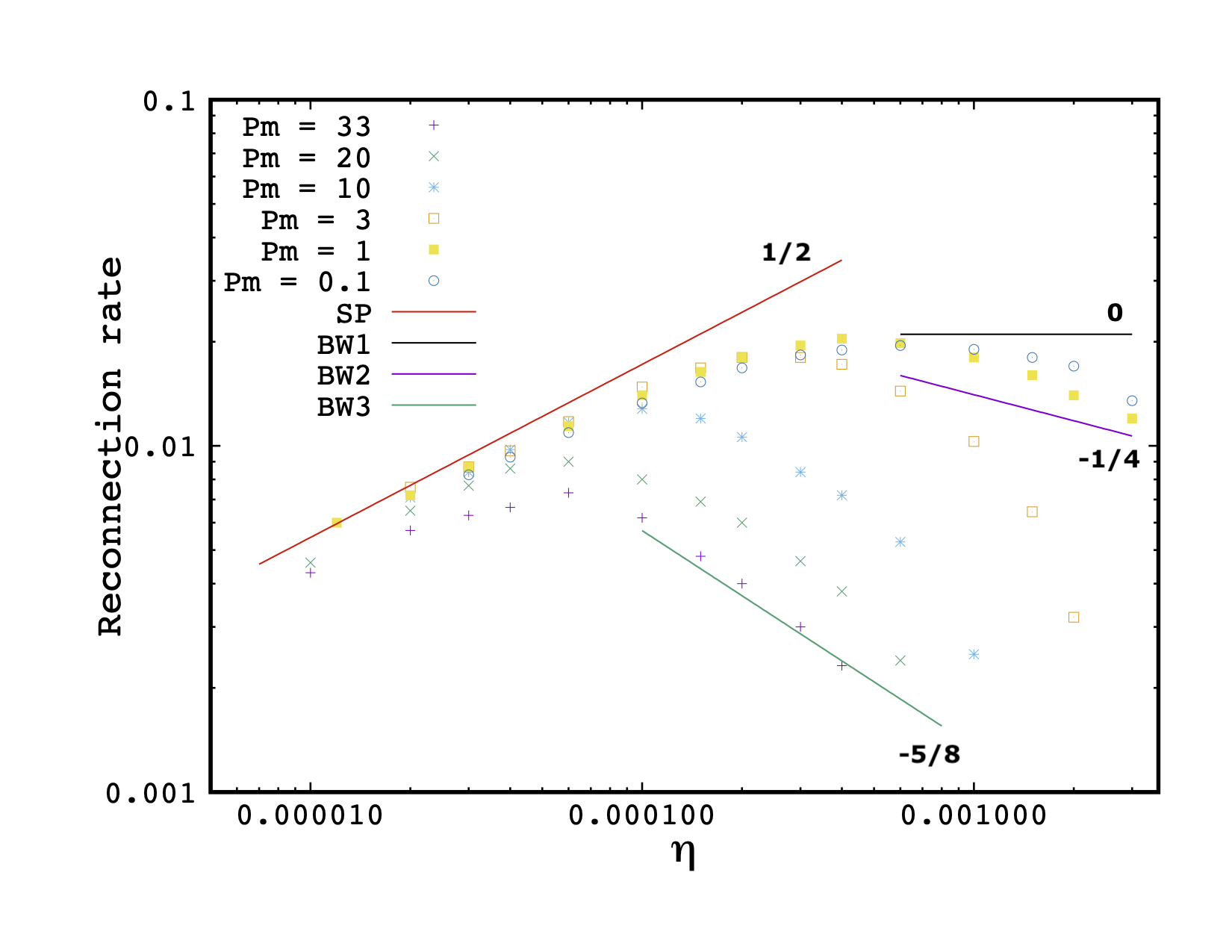}
    \caption{Reconnection rate estimated from the peak density current via $\eta J_M$ for the coalescence setup, corresponding to the cases shown in the previous figure. The resulting SP
    (the fitted law is $1.7 \times  \eta^{1/2}$) and BW1-2-3 expected scaling laws in $\eta^{0} - \eta^{-1/4} - \eta^{-5/8}$ are also plotted for comparison.
     }
   \label{fig7}
\end{figure}

 \begin{figure}
     \centering
 \includegraphics[scale=0.32]{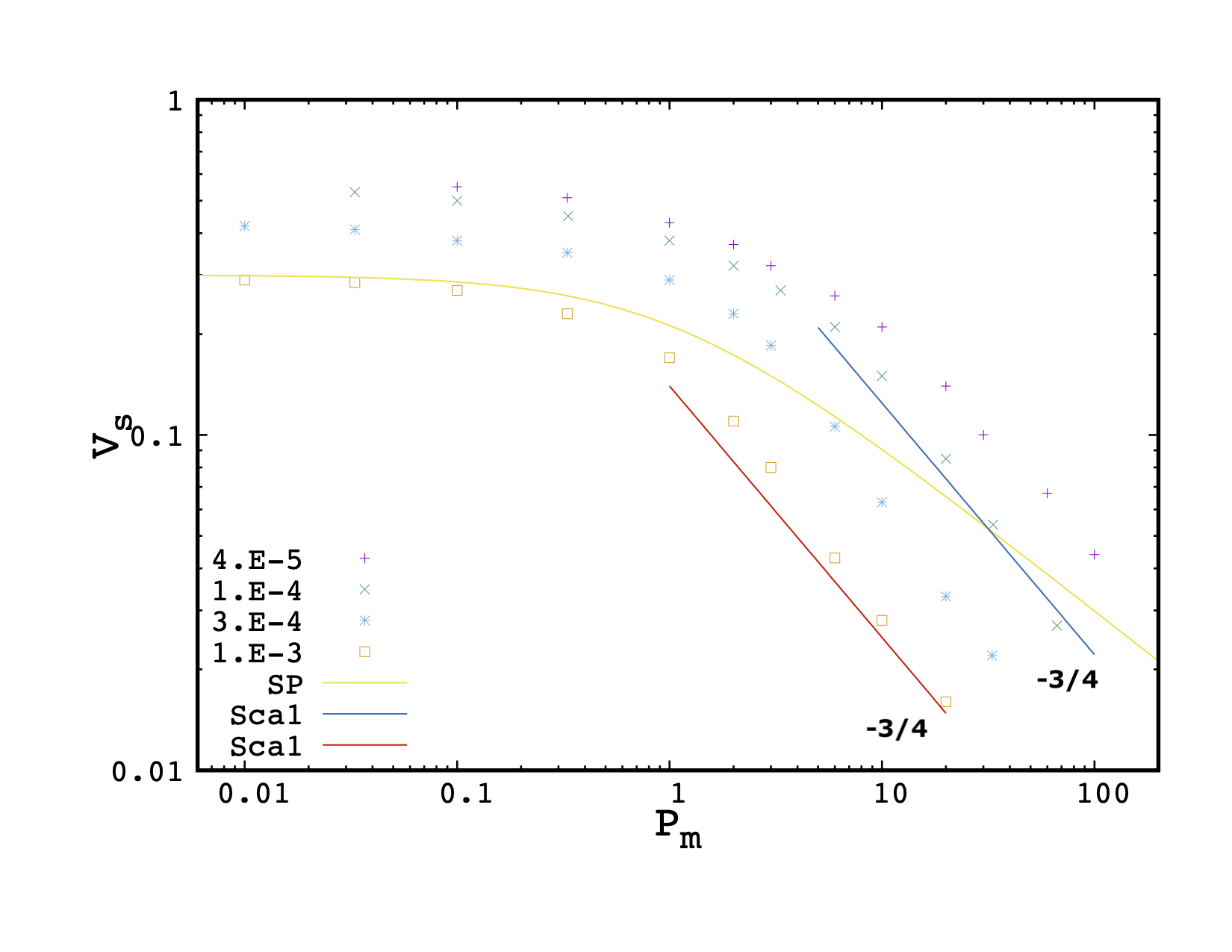}
 \includegraphics[scale=0.32]{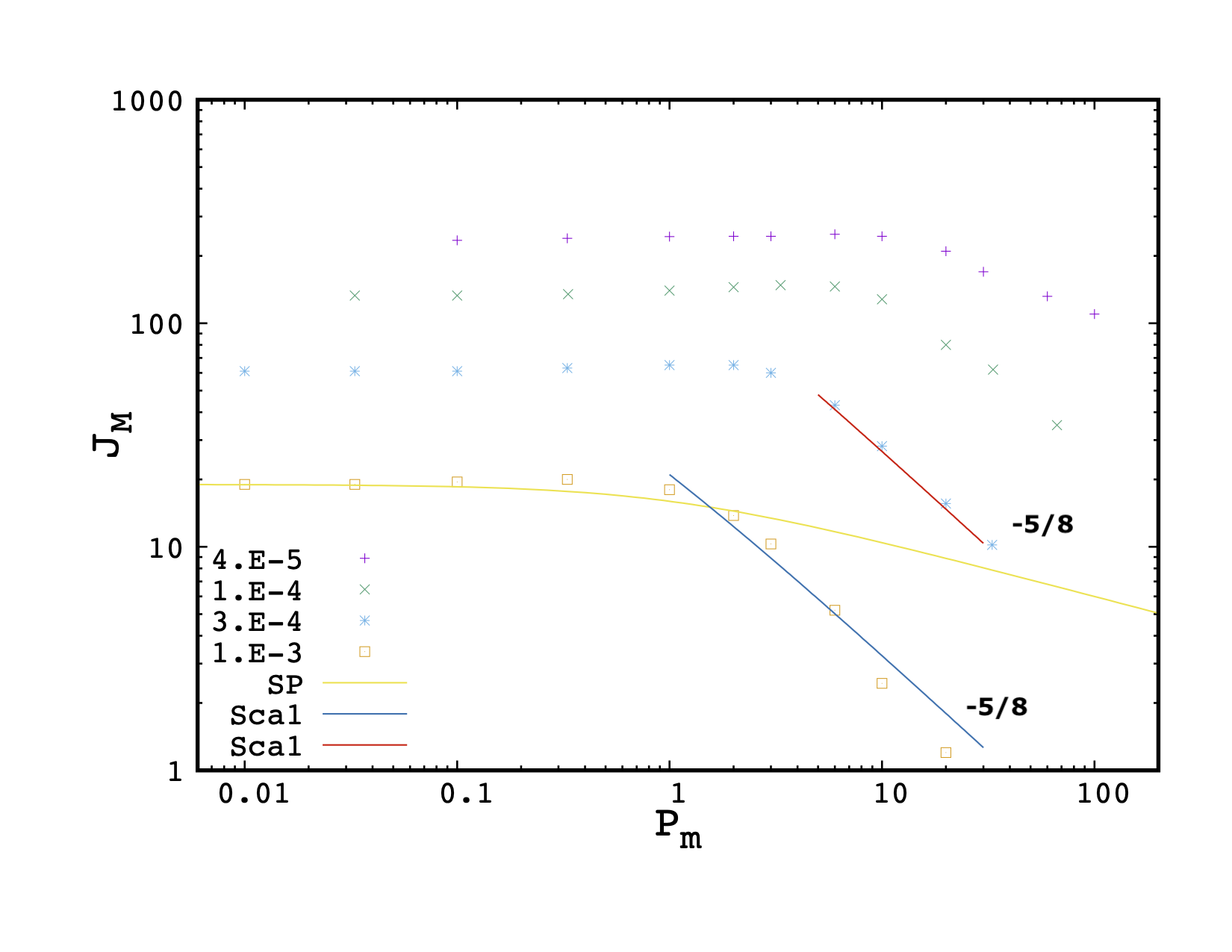}
    \caption{Scaling study of different parameters ($V_s$, and $J_M$) with magnetic Prandtl parameter $P_m$ deduced for runs of the coalescence setup at four different resistivities
    (i.e $\eta = 4 \times 10^{-5},  1 \times 10^{-4},  3 \times 10^{-4}$, and $1 \times 10^{-3}$). The SP expected scalings in $(1 + P_m)^{-1/2} $ and 
     $(1 + P_m)^{-1/4}$ for $V_s$ and $J_M$ (see Section 2) respectively are also plotted for comparison. Additional power laws (Sca1) following $P_m^{-3/4} $
     and  $P_m^{-5/8}$ dependences  for $V_s$ and $J_M$ plots respectively are also plotted.
     }
   \label{fig8}
\end{figure}

\subsection{Magnetic reconnection associated to the coalescence instability}
The ideal MHD stability of the coalescence setup has been examined by \citet{bon83} with a reduced linear MHD framework. The use of the minimum energy principle
shows that it is unstable due to the current-driven term $\frac {d J_e} {d \psi_e} > 0 $ that makes the second order variation of the associated potential energy negative.
Moreover, the resulting linear growth rate scales as $\epsilon^{3/4}$. 

An overview of the time evolution of the system is plotted in Figure 4 with snapshots of the current density overlaid by a few magnetic field lines taken at different times.
This corresponds to a simulation obtained with FINMHD, where a run employing $\epsilon = 0.2$ 
and $\alpha = \pi/2$ in a periodic domain along the $x$ direction $[-1: 1]$ is considered, i.e. with $k = 2 \pi$. Indeed, a current sheet is observed to form at $ t \simeq 15 t_A$ (localized at $x = 0.5$), leading
to a magnetic reconnection process between the two islands which ends up at $t \simeq 20 t_A$. The Alfv\'en time $t_A$ is defined as $t_A = L_c/V_A$, with $L_c$ the characteristic unit
length (here the half periodic length of the Fadeev configuration) and $V_A = 1$ (i.e. $B_0 = 1$) in our units.
The maximum vorticity $\Omega_M$ and maximum current density taken at the $X$-point of the current sheet
$J_M$ are measured as function of time and are plotted in Figure 5 for the same run. The linear phase is clearly visible with the vorticity variation leading to an estimate of the
linear growth rate $\gamma t_A \simeq 0.1$ that is in agreement with values reported in the literature (see \citet{kno06} and references therein). Note that, the numerical noise is sufficient to let the system
develop the unstable mode, without the help of any small perturbation added initially, as stated in the previous section.
During the reconnection phase, the maximum current density $J_M$
is not constant, as it increases, reaching a peak value before decreasing. Consequently, this complicates the comparison with Sweet-Parker model which assumes a
steady-state reconnection. However, as done previously, we can use the peak
values of $\Omega_M$, and $J_M$ in order to test the dependence with the dissipation parameters ($\eta$ and $\nu$) in the following study. For the run illustrated in Figures 4-5,
this corresponds to a time $t \simeq 16.5 t_A$. Note that for the lowest dissipation parameters used in this study, typically when $\eta$, and/or $\nu$ are of the order $10^{-5}$,
this first peak value is followed by other secondary peaks of lowest amplitude because of a sloshing effect described by \citet{kno06}. We thus consider only the first peak
for the scaling laws (see below). 

As a first testing procedure of the theoretical scaling laws previously obtained in this work (see Section 2), we examine the dependence
of the important parameters with the resistivity parameter $\eta$ for a fixed value
of the magnetic Reynolds number $P_m$. The results are plotted in Figure 6 for different $P_m$ values, i.e. $P_m = 0.1, 1, 3, 10, 20$, and $33$.
The outflow velocity $V_s$ is shown to become independent of $\eta$ in the limit of very small resistivity values, as expected from SP model. When $P_m$
is additionally much lower than unity (case $P_m = 0.1$), it is exactly the Alfv\'en speed that is expected from the SP model with a value $0.65$ in our units. This latter value is in
very good agreement with the value reported by \citet{kno06} (see Figure 7). This saturated $V_s$ value depends on $P_m$ (see dependence with $P_m$ in the second testing procedure below).
Moreover, in the opposing limit of relatively high
$\eta$ values, one can see a decreasing dependence with the resistivity that we approximate to be exponential-like $ \eta^{- \alpha}$ with $\alpha$ a real positive exponent.
We also obtain that $\alpha$ is not constant as it typically varies between $\alpha  \simeq 1/3$ (for $P_m = 0.1$), $\alpha  \simeq 1/2$ (for $P_m = 1$), and $\alpha  \simeq 3/4$ (for $P_m = 33$).
As the outflow velocity depends on $B_e$ and $L$ as obtained in Section 2, this reflects the dependence of these two parameters (taken as fixed and constant in the standard SP model).
A dependence of $B_e$ in $\eta^{-1/2}$ was proposed by \citet{delu92} in order to explain numerical results obtained by \citet{bisk80} (BM study hereafter) where $J_M$
scales as $\eta^{- 1}$ instead of $\eta^{- 1/2}$ (i.e. the SP value). We thus call these three dependences BW1 (case with $\alpha = 1/3$), BW2 ($\alpha = 1/2$), and BW3 ($\alpha = 3/4$).
Looking at the corresponding maximum peak values for the current density $J_M$, one can see the transition between the SP scaling law $\eta^{- 1/2}$ (for small
resistivity limit) and scaling laws in $\eta^{- 1}, \eta^{- 5/4}, \eta^{- 13/8}$ for BM1, BM2, and BM3 regimes respectively. This follows from the $B_e^{3/2}$ parameter for $J_M$
that consequently scales as $\eta^{- 1/2 - 3 \alpha/2}$ (see Section 2). A fitted law for $J_M$ in the small $\eta$/$P_m$ limits is $J_M = 1.7 \times \eta^{- 1/2 }$ in our units, leading to 
the estimate for the length $L  \simeq 0.1$ as $B_e \simeq 0.65$, that is in agreement with the value deduced from direct estimate from our simulations.
A similar conclusion can be drawn from  the peak vorticity $\Omega_M$.
These above results lead to the reconnection rate $\eta J_M$ plotted in Figure 7. Thus, the SP reconnection rate scaling law in $\eta^{1/2}$ is obtained in the small
resistivity limit, while in the opposite limit scaling laws in $\eta^{0}$, $\eta^{-1/4}$, and $\eta^{-5/8}$ are checked for BM1, BM2, and BM3 regimes respectively.
Finally, note that we have found that the transition between the SP-like scaling and different BW-like ones are mainly due to the $B_e$ dependence with the
resistivity coefficient $\eta$. There is also another effect due to the dependence of the length $L$ with $\eta$. This is visible for the very highest $\eta$ values employed
in the simulations in Figure 6. This latter effect is however found to be weaker compared to the $B_e$ one reported above and more difficult to explore into detail.

Additionally, the previous figures also clearly show that the results depend on $P_m$ in the vanishing $\eta$ limit. For example, the outflow velocity $V_s$ is $0.65$ for 
small $P_m$ (i.e. $P_m = 0.1$), and $V_s = 0.13$ for $P_m = 33$. Thus, in a second testing procedure, we examine the dependence of the important parameters
with the magnetic Prandtl number for different fixed values of the resistivity parameter $\eta$. Typically, we use $\eta = 4 \times 10^{-5},  1 \times 10^{-4},  3 \times 10^{-4}$,
and $1 \times 10^{-3}$.
 The results are plotted in Figure 8 for $V_s$ and $J_M$. A transition between a SP-like scaling law obtained low low $P_m$ values (typically lower than unity) and a 'Sca1' power law
in $P_m^{-3/4}$ for $V_s$, and in $P_m^{-5/8}$ for $J_M$. The results for $\Omega_M$ (not shown) are similar to the $J_M$ plot. According to the analytical scalings deduced in
Section 2, this infers a dependence for $B_e$ in $P_m^{-1/4}$ for large enough $P_m$ values. This is also valid for the four resistivity values investigated in this analysis.

In summary, we have obtained that the standard SP scaling laws are recovered only in the small $\eta$ and $P_m$ limit as expected from SP model. When considering
the opposite limit, modified scalings must be considered mainly because the inflowing magnetic field $B_e$ is not constant and depends on these two dissipative parameters.
More precisely, a dependence $B_e \propto \eta^{-\alpha}$ is deduced from our simulations, with $\alpha$ varying in the range $[1/3 : 3/4]$ when $P_m$ is varying
between small values (i.e. much lower than unity) and large values (i.e. much larger than unity). This is similar to a scaling deduced by Biskamp and Welter, where
$B_e \propto \eta^{-1/2}$ was previously reported in an incompressible numerical study. The main consequence of this dependence is to alter the SP scalings because
of an additional $B_e^{3/2} \propto \eta^{-3\alpha/2}$ factor in the peak density current $J_M$ and associated reconnection rate (see Figure 7 for example). Moreover,
a dependence $B_e \propto P_m^{-1/4}$ is deduced when $P_m$ is not small enough (typically when $P_m  >> 1$), whatever the resistivity value. One must note the smallest
resistivity value employed in this coalescence setup is $\eta \simeq 10^{-5}$, as it corresponds to a Lundquist number close to the critical value for plasmoid regime.
When writing this paper, we were aware of a recent study using a very efficient MHD finite-element code similar to FINMHD, which reports a $J_M$ value for
the peak current density of $140$ and outflow speed of $0.38$ for a run with $\eta = 1  \times 10^{-4}$ and $P_m = 1$ for the same coalescence setup \citep{tan21}. These values are exactly the values
we deduced from our simulations (see Figure 6), building thus strong confidence in our results.

 \newpage

\subsection{Magnetic reconnection associated to the tilt instability setup}
The ideal MHD stability of the tilt mode has been examined by \citet{ri90}. The energy principle in the reduced MHD approximation shows
that the equilibrium is unstable with a linear eigenfunction that is a combination of rotation and outward displacement. The current-driven term is again
at the origin of this ideal mode. This result has been numerically
confirmed in reduced MHD framework \citep{lan07} and using full compressible MHD \citep{kep14, rip17}. The instability called the tilt mode proceeds with
a linear growth rate $\gamma t_A \simeq 1.3-1.4$, where the Alfv\'en time is now $t_A = R/V_A$ (with $R$ being the dipole radius and $V_A$ being defined with the asymptotic field $B_0$).

An overview of the time evolution of the system is plotted in Figure 9 with snapshots of the current density overlaid by a few magnetic field lines taken at different times.
This corresponds to a simulation obtained with FINMHD, where a run employing $\eta = \nu = 1 \times 10^{-3}$ is chosen.
The maximum vorticity $\Omega_M$ and maximum current density $J_M$ taken over the whole domain measured as function of time are plotted in Figure 10 for the same run. 
The reconnection phase is starting when two twins curved curent layers (of opposite sign) are formed, typically at the frontier between the two closed field lines regions and
the external region (see second snapshot in Figure 9). Then, closed filed lines reconnect with open ones, leading to new field lines (see the third snapshot).
As an important result, and contrary to the coalescence mode, the reconnection phase is proceeding with a nearly constant current density (that is also close to the value
for the first peak) for the tilt case, as one can see in Figure 10.

 \begin{figure}
     \centering
 \includegraphics[scale=0.16]{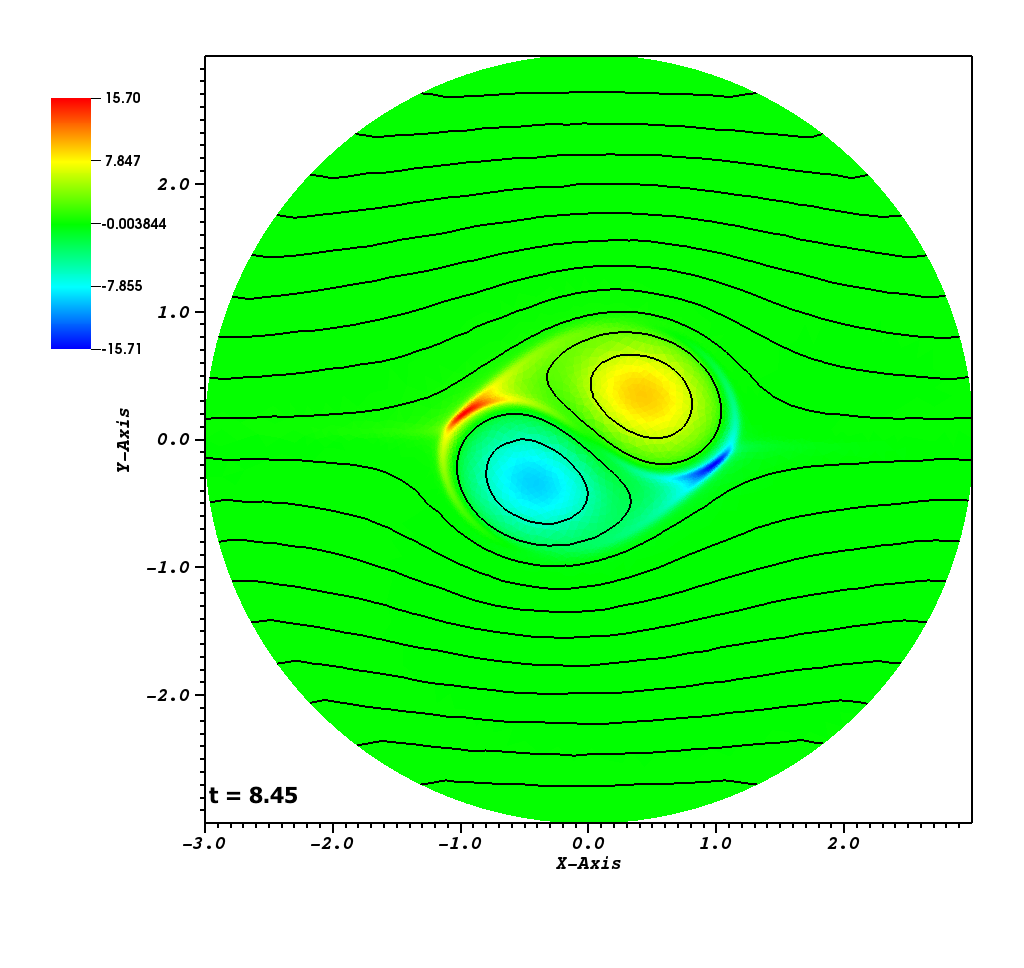}
 \includegraphics[scale=0.16]{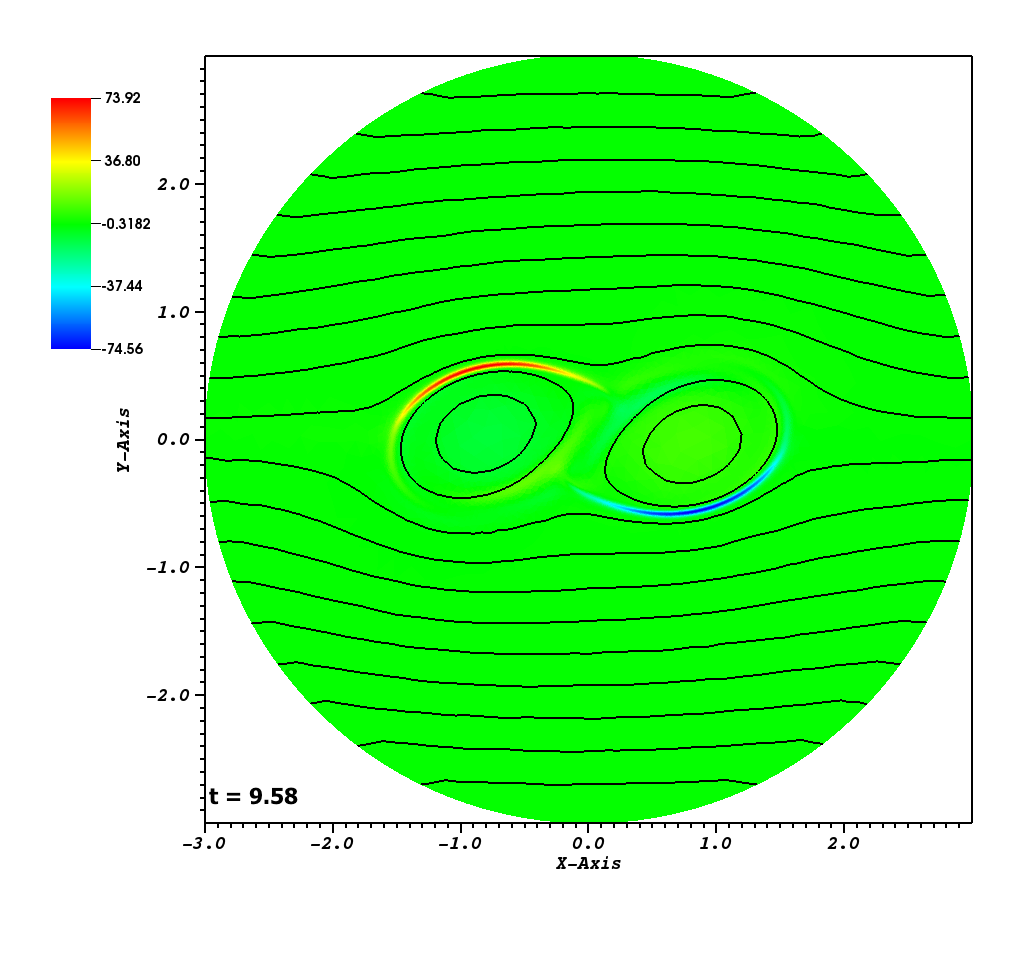}
 \includegraphics[scale=0.16]{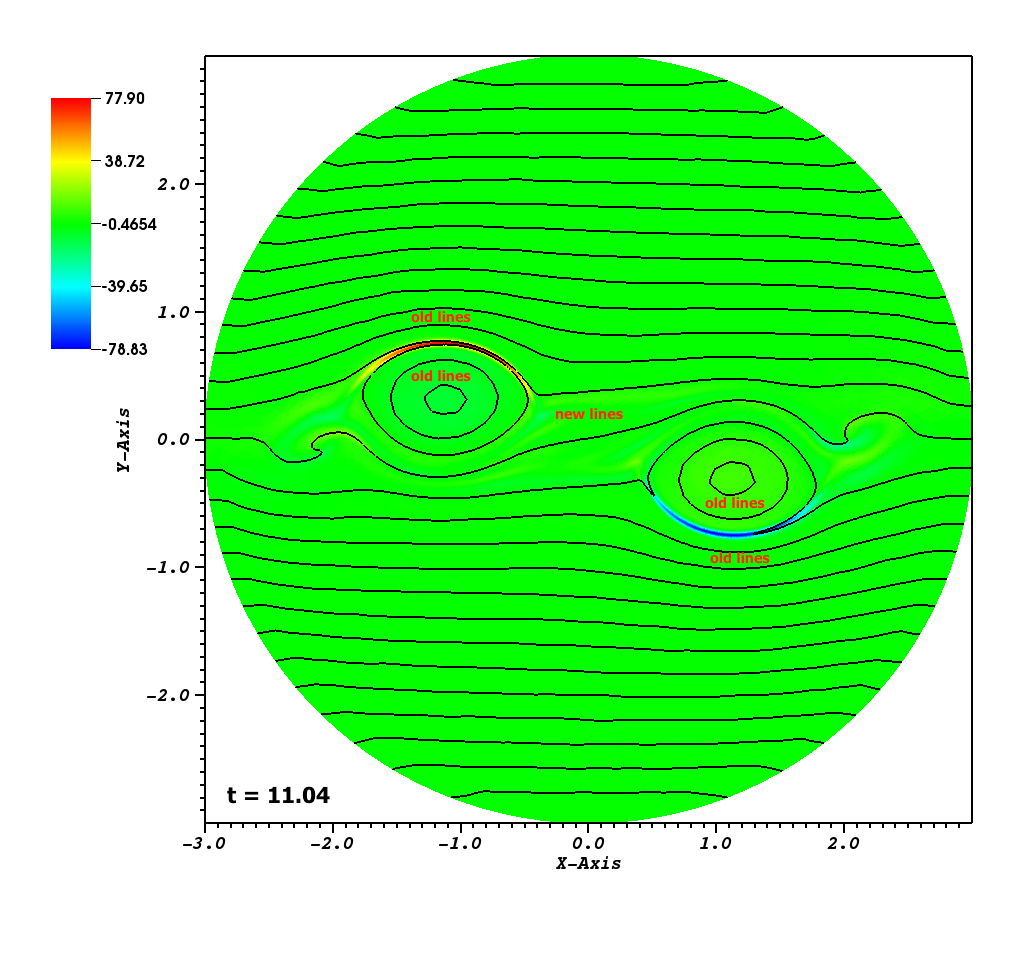}
      \caption{Snapshots taken at different times of the current density (colored iso-contours) overlaid with magnetic field lines. The run is obtained for
    the tilt setup using $\eta = \nu = 1  \times 10^{-3}$.
     }
   \label{fig9}
\end{figure}

    \begin{figure}
     \centering
 \includegraphics[scale=0.3]{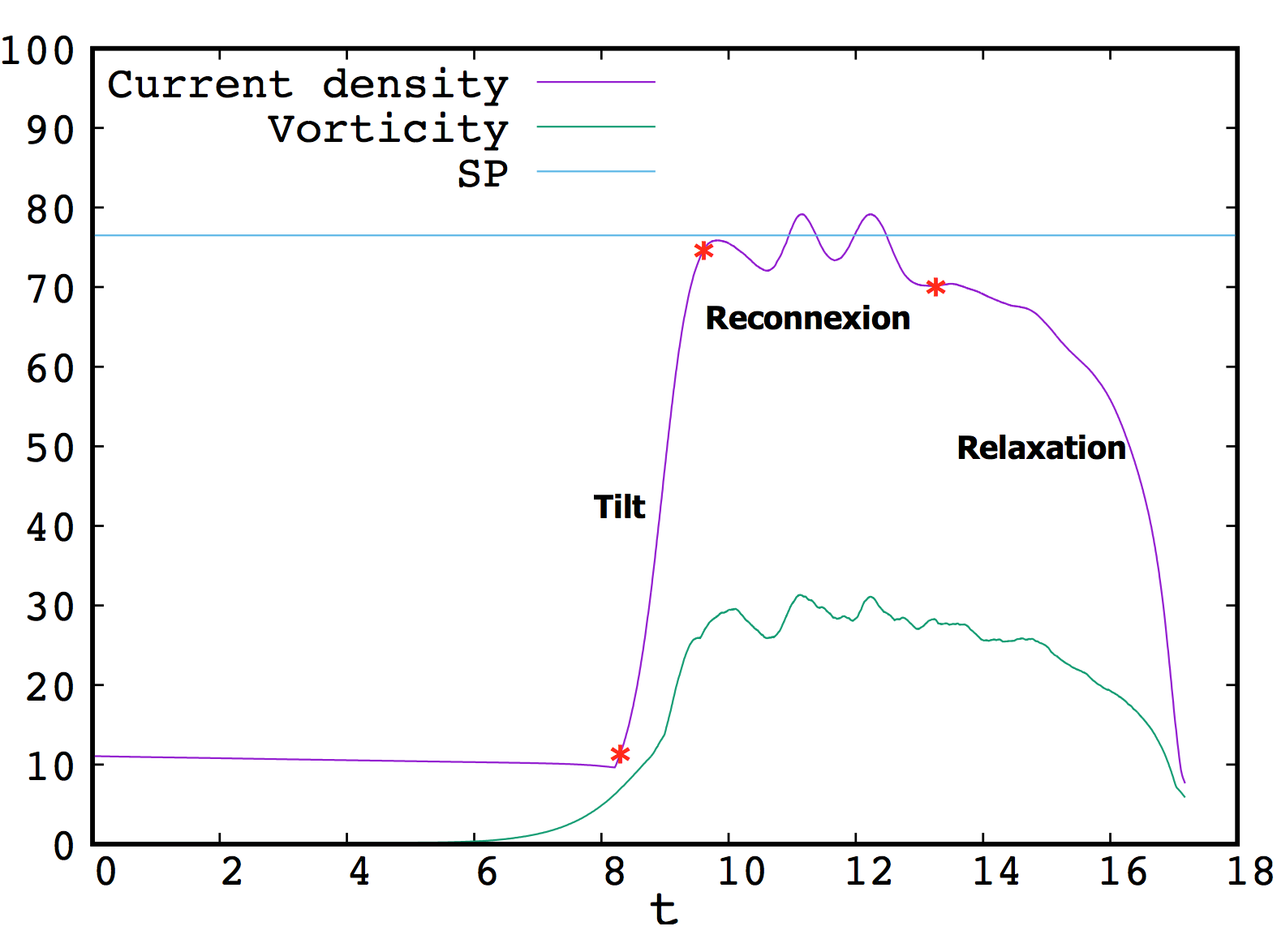}
    \caption{Maximum vorticity $\Omega_M$ and maximum current density $J_M$ as a function of time
    fo the run corresponding to the previous figure. An horizonthal line indicates the average current value (i.e. $77$) during reconnection phase,
     which also agrees with the value predicted from the SP theory.
        }
  \label{fig10}
\end{figure}

Following the same procedure as taken for the coalescence setup, we first investigate the dependence of the important parameters with the resistivity coefficient $\eta$ for different fixed values
of the magnetic Reynolds number $P_m$ in the range $[0.1 : 100]$.
The results are illustrated in Figure 11. First, one see that the deviation from SP scaling observed in $J_M$ and $\Omega_M$ curves for the highest
values of the resistivity are weaker than the one obtained from the coalescence setup. Indeed, a modified scaling law in $\eta^{-3/4}$ is obtained for the largest $P_m$ cases. This is
in agreement with the results for $V_s$ showing a transition towards a dependence that scales approximatively like $\eta^{-1/6}$, in agreement with a dependence for the inflowing
magnetic field $B_e \propto \eta^{-1/6}$ for the tilt setup, according to scaling laws of Section 2 where we neglect the dependence of $L$ with $\eta$. This dependence is even weaker for the smallest $P_m$ values.
As concerns the reconnection rate, estimated via the term $\eta J_M$, the results plotted in Figure 12 remains close to the SP scaling that follows a law as $2.8 \times  \eta^{1/2}$ in ou units
(for vanishing dissipative parameters). This agrees well with the inflowing value of the magnetic field $B_e \simeq 1.88$ (see Figure 11) and the half-length estimate $L \simeq 1$, for vanishing
$\eta$ and $P_m$. 

 \begin{figure}
     \centering
 \includegraphics[scale=0.4]{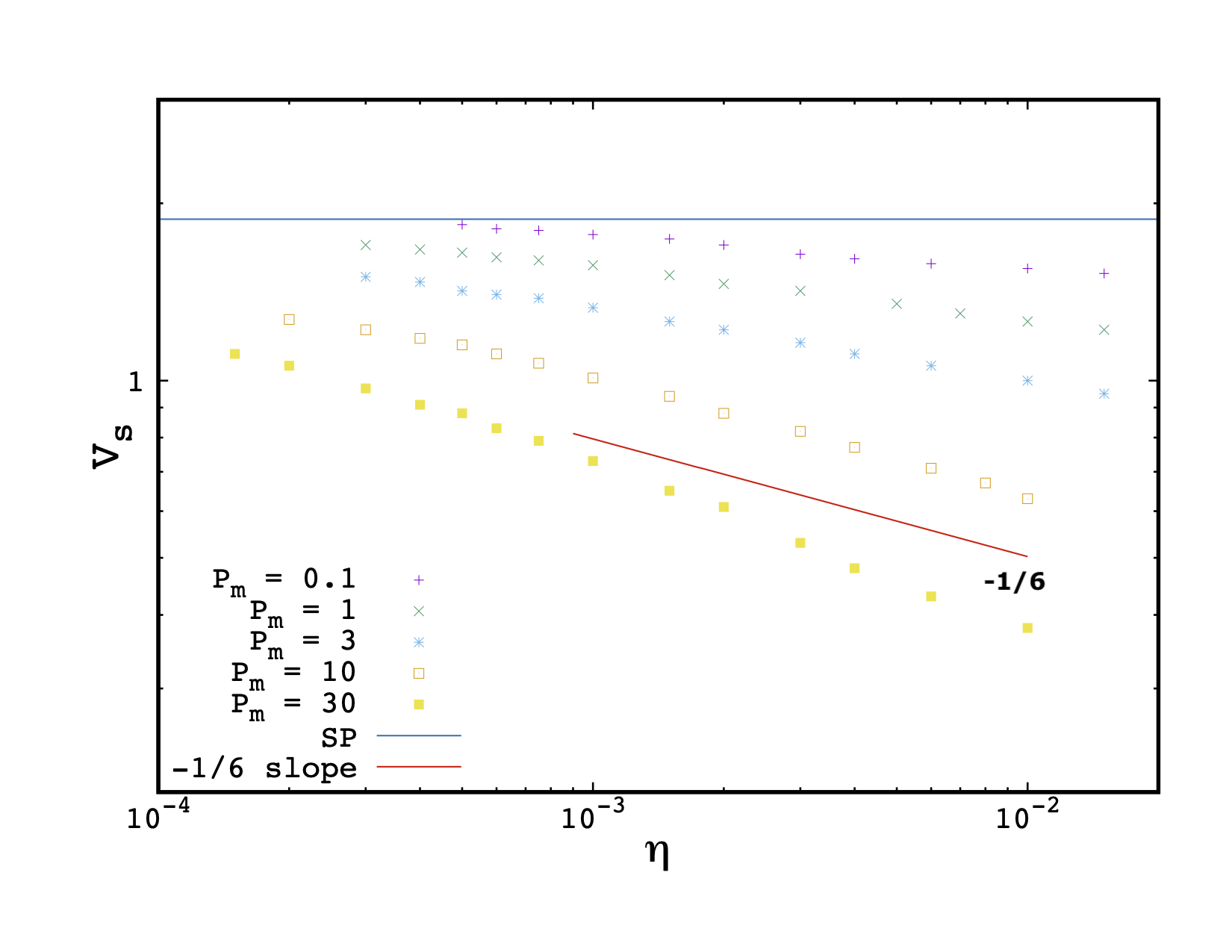}
 \includegraphics[scale=0.32]{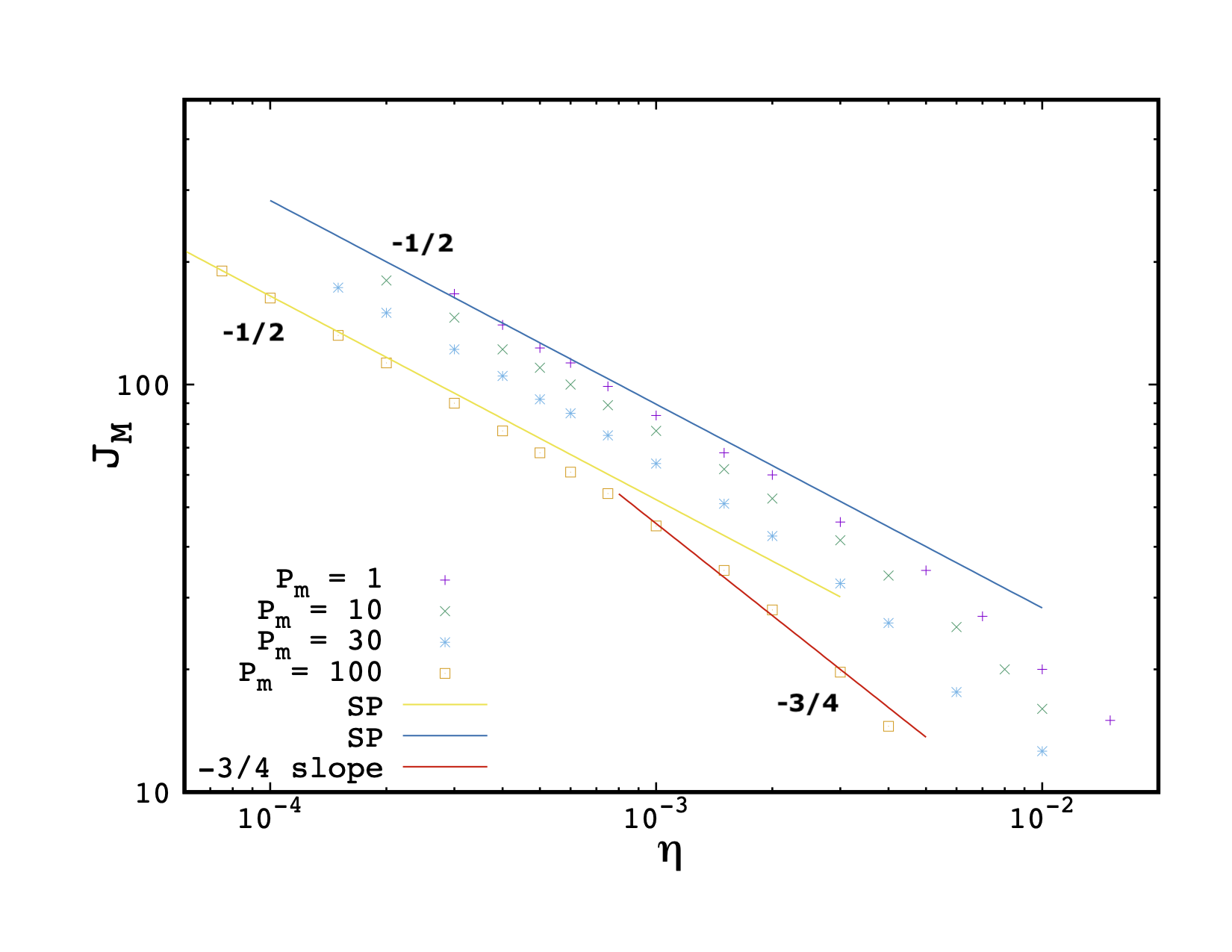}
 \includegraphics[scale=0.32]{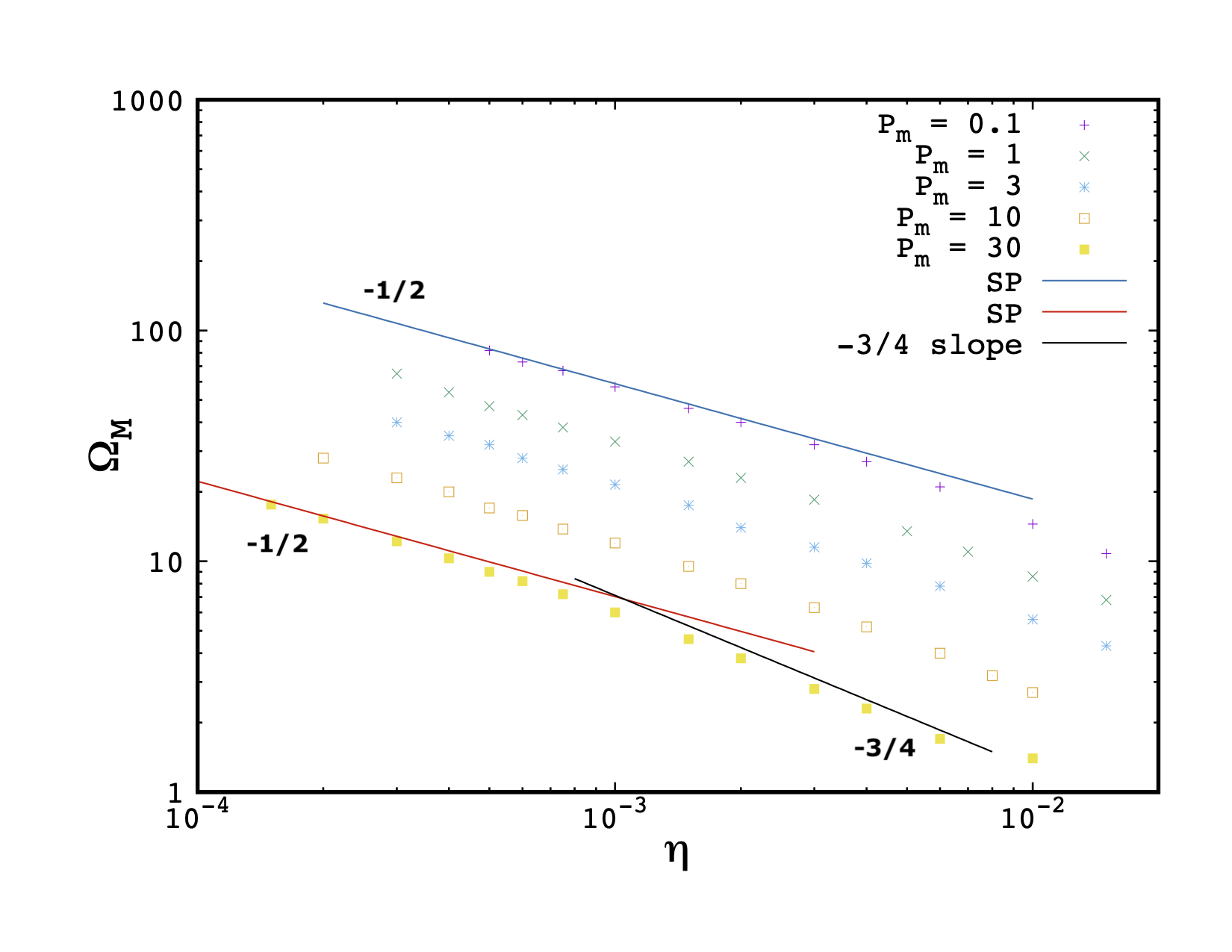}
      \caption{Scaling study of different parameters ($V_s$, $J_M$, and $\Omega_M$) with resistivity parameter $\eta$ deduced for many runs of
    the tilt setup at different magnetic Prandtl $P_m$, Expected SP scalings, and power laws with exponents of $-1/6$ and $-3/4$ are also plotted for comparison.}
   \label{fig11}
\end{figure}

 \begin{figure}
     \centering
 \includegraphics[scale=0.4]{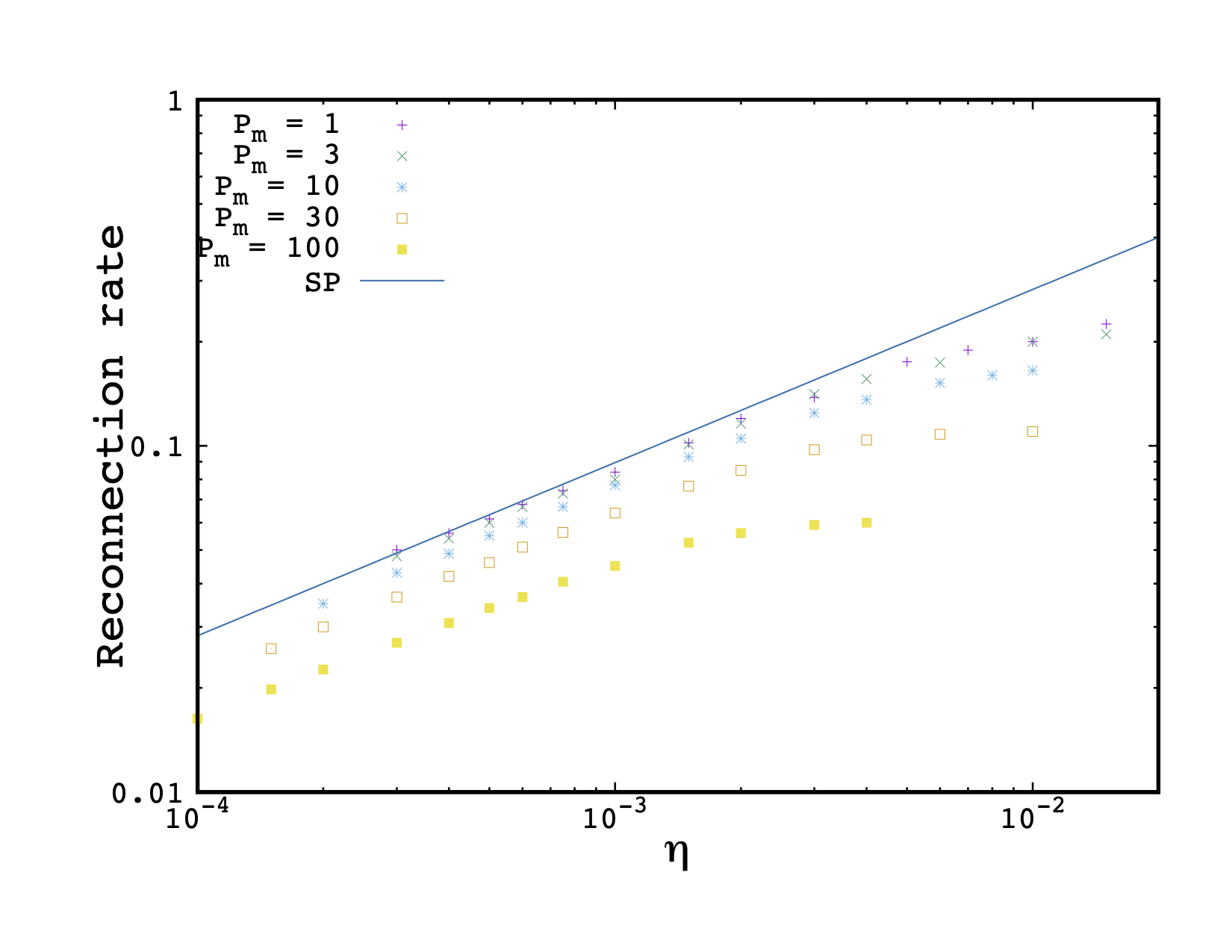}
    \caption{Reconnection rate estimated from the maximum density current via $\eta J_M$ for the tilt setup, corresponding to the cases shown in the previous figure. The resulting SP
    (the fitted law is $2.8 \times  \eta^{1/2}$) valid for the small resistivity and Prandtl number limit is also plotted for comparison.
     }
   \label{fig12}
\end{figure}

As done for the coalescence setup, In our second testing procedure, we examine the dependence of the important parameters
with the magnetic Prandtl number for different fixed values of the resistivity parameter $\eta$. Typically, we use $\eta = 3 \times 10^{-4},  5 \times 10^{-4},  2 \times 10^{-3}$,
and $ 6 \times 10^{-3}$. The results are plotted in Figure 13. First, the expected SP scaling for $V_s$ with the $(1 + P_m)^{-1/2}$ dependence is retrieved at small enough $P_m$.
However, a transition to a power law in $P_m ^{-0.25} - P_m^{-0.4}$ (the case  $P_m^{-0.25}$ being for the smallest resistivity value)
is obtained in the high $P_m$ limit. Thus, the slope becomes weaker at large $P_m$ for the tilt setup, that is the opposite effect compared to the results obtained for the coalescence setup
where the transition shows a slope increase with a power law in $P_m^{-0.75}$ for $V_s$ (see Figure 8). As a consequence, assuming that $L$ is independent of
$P_m$, this infers now a weak dependence for the magnetic field in this high $P_m$ limit, $B_e  \propto  P_m^\beta$ with a positive exponent $\beta \simeq 0.1-0.25$.

In summary, we have obtained that the standard SP scaling laws are recovered only in the small $\eta$ and $P_m$ limit as expected. When considering
the opposite limit, modified scalings must be considered mainly because the inflowing magnetic field $B_e$ is not constant and depends on these two dissipative parameters.
More precisely, a dependence $B_e \propto \eta^{-1/6}$ is deduced from our simulations in the opposite high resistivity regime, independently of the magnetic Prandtl number $P_m$.
Moreover, $B_e  \propto  P_m^\beta$ with a positive exponent $\beta \simeq 0.1-0.25$ that slightly depends on the resistivity value in the large $P_m$ limit.
As for the coalescence instability, the smallest resistivity value employed for the tilt setup is $\eta \simeq 1 \times 10^{-4}$ roughly coinciding with the critical Lundquist number
for plasmoid instability.

\section{Conclusion}

In this study, we have revisited the well known Sweet-Parker model for 2D incompressible magnetic reconnection, focussing on the possible
extra dependences of the length $2L$ and inflowing magnetic field $B_e$ with the dissipation parameters taken to be $\eta$ and $P_m$.

Taking two different setups involving unstable ideal MHD equilibria (namely the coalescence and tilt modes) to form the current sheet, we
have illustrated the effect of $B_e (\eta, P_m)$ using numerical simulations.
As expected, the standard visco-resistive SP scaling is retrieved in the limit of small enough resistivity and magnetic
Prandtl number values. However, non negligible deviations are observed in the other limit. More precisely, a first dependence $B_e \propto \eta^{-\alpha}$ is
deduced, with the parameter $\alpha \simeq 1/6$ for the tilt, and $\alpha$ varying in the range $[1/3 : 3/4]$ for the coalescence mode. The second
dependence observed is $B_e  \propto  P_m^\beta$, with the parameter being positive as $\beta \simeq 0.1-0.25$ for the tilt, and negative $\beta \simeq - 0.25$
for coalescence. The deviations are thus weaker for the tilt instability setup when compared to the coalescence one. Consequently, the initial unstable configuration
is also important to this respect. This is not surprising, as for example the geometrical structure of the current sheet is evidently different during the tilt instability
(e.g. curved current layer) when compared to the coalescence setup.

We hope that this study will be useful in order to help to correctly interpret the results of numerical simulations involving magnetic reconnection. It
emphasizes the importance of determining the dissipation parameters like the resistivity and viscosity, that are not always explicitly known. This is
for example the case when they are dominated by truncation errors due to the numerical scheme discretization. This is true for testing procedures using MHD codes
in the SP regime. This is also the case when one to focus on the plasmoid-dominated regime (i.e. for very small resistivity values or equivalently very high $S$ values)
without considering the viscosity effect via the magnetic Prandtl number $P_m$. Indeed, as one can see in Figures 7 and 12, the reconnection rate is even
noticeably affected for $P_m$ values of order $10$ for the smallest resistivity values employed in this work. 

Magnetic reconnection is believed to be the underlying mechanism that explains explosive events observed in many magnetically dominated plasmas. This is for example the case for flares in the solar corona.
However, the timescales involved in classical two-dimensional (2D) reconnection models within the macroscopic MHD regime are too slow to match the observations or experiments.
Indeed, the reconnection rate predicted by Sweet-Parker model  is too low by a few (or even many) orders of magnitude for the relevant Lundquist numbers.
For typical parameters representative of the solar corona, $S$ is of order $10^{12}$, 
leading to a normalized SP reconnection rate of order $10^{-6}$ at negligible viscosity, that is much lower than the value of $10^{-2}- 10^{-1}$ required to match the observations. However, the
plasmoid regime that is relevant at such huge value of  the Lundquist number,  is a stochastic time-dependent reconnection solution with a fast time-averaged rate independent of $S$.
The normalized reconnection rate values reported in the literature are of order  $0.01$, much higher than the Sweet-Parker rate, and thus could be sufficient. However, in such studies the viscosity effect is often neglected,
mainly for the sake of simplicity. The (collisional) viscosity parameter value is expected to be at least equal to the resistivity one in the solar corona.
Future studies including the $P_m$ effect are thus required to explore this regime.

 \begin{figure}
     \centering
 \includegraphics[scale=0.33]{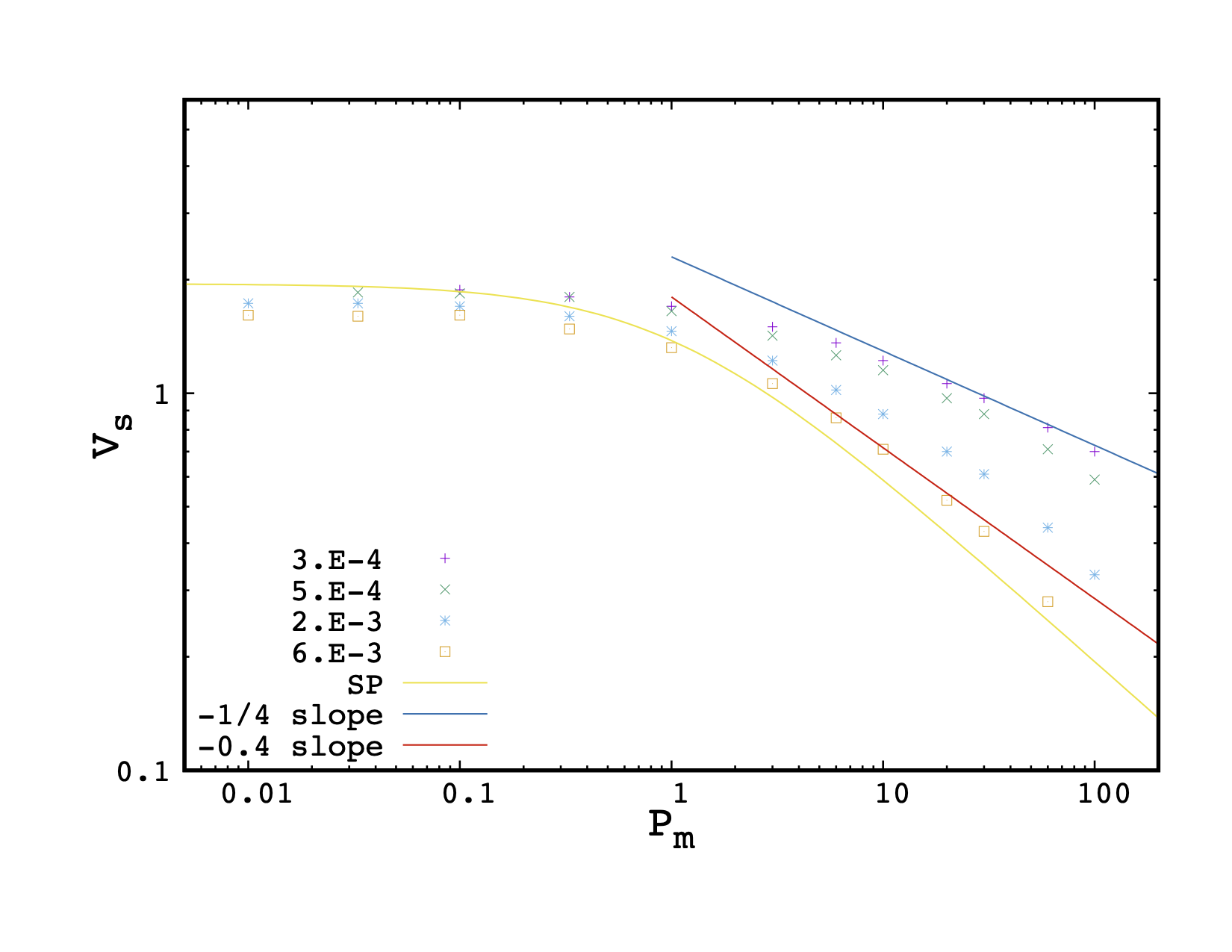}
 \includegraphics[scale=0.33]{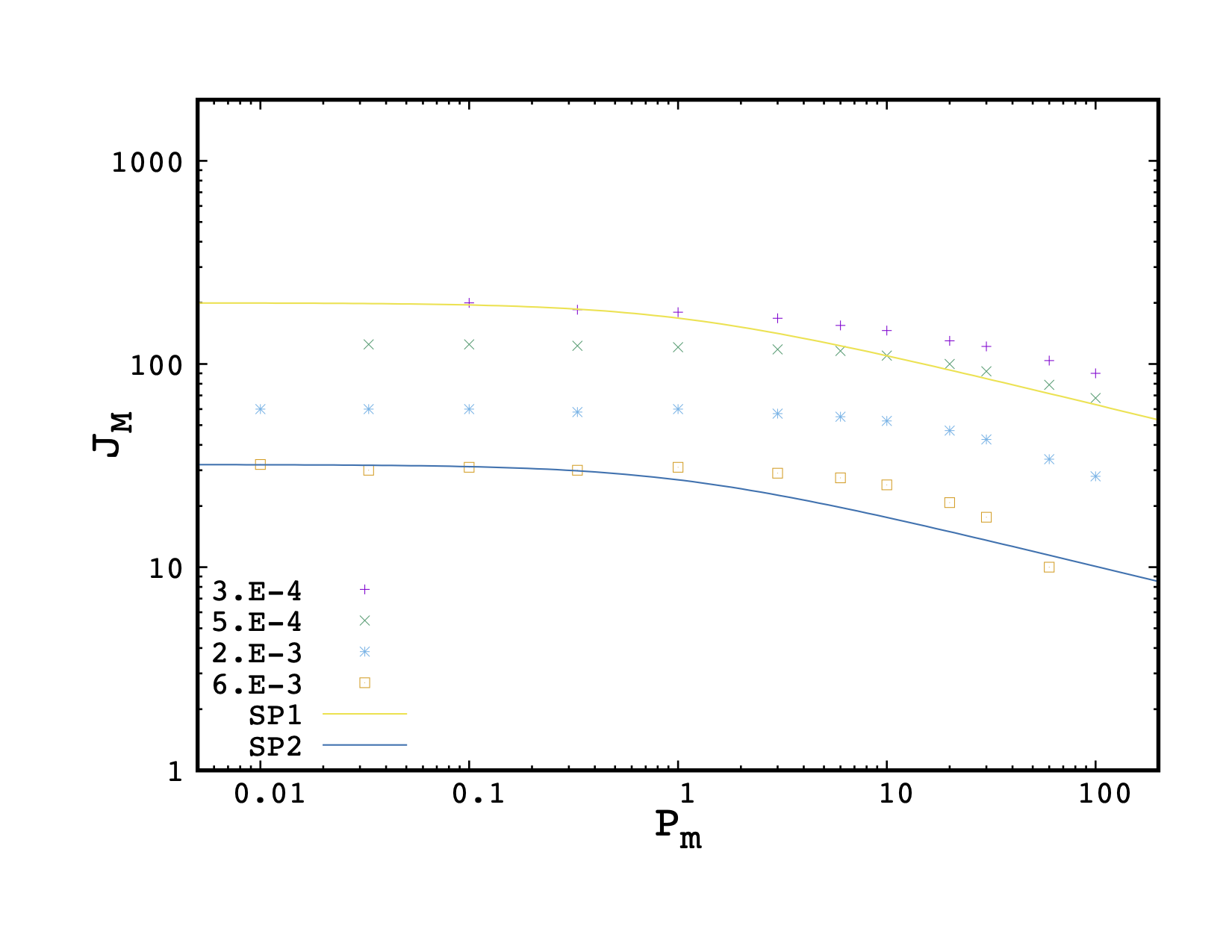}
 \includegraphics[scale=0.33]{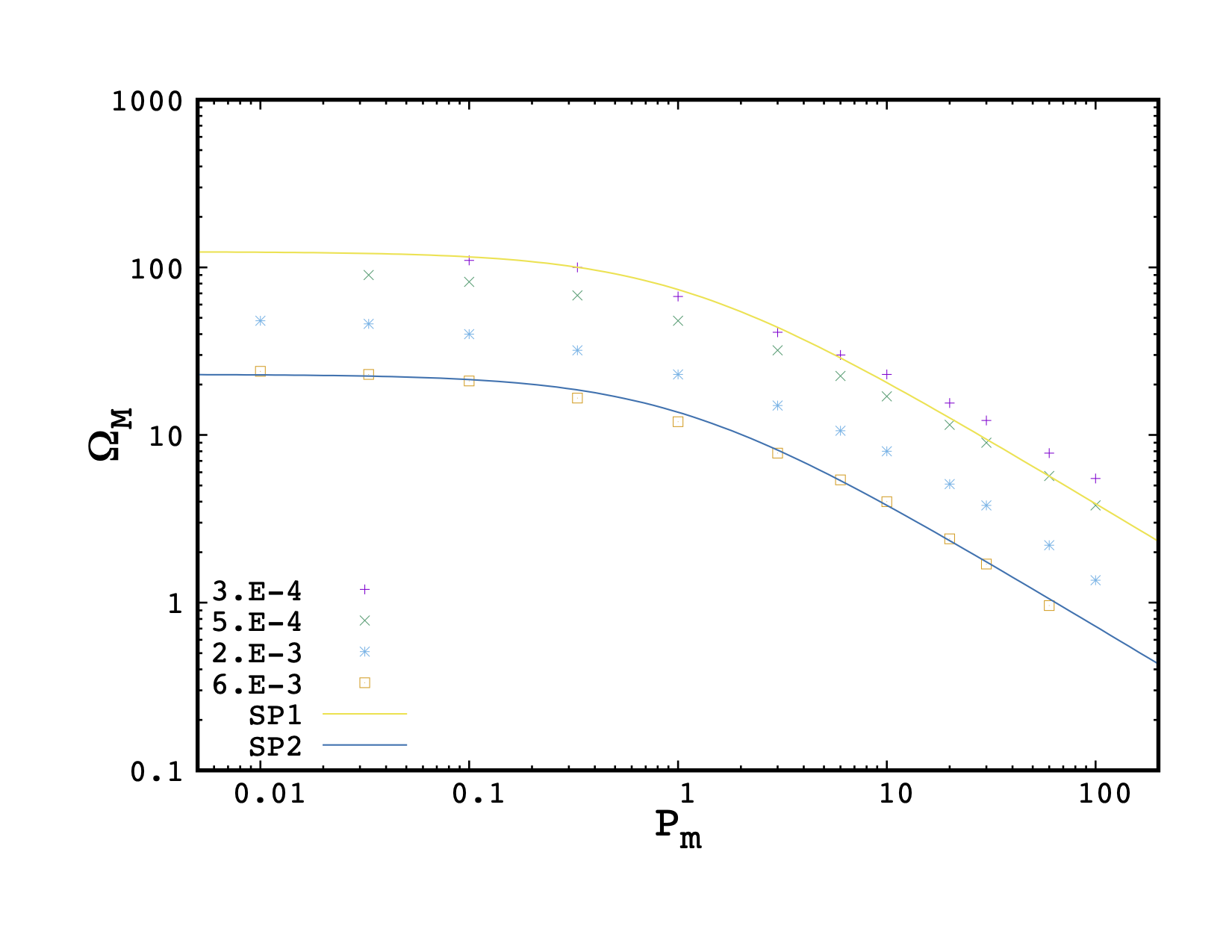}
    \caption{Scaling study of different parameters ($V_s$, $J_M$, $\Omega_M$) with magnetic Prandtl parameter $P_m$ deduced for runs of the tilt setup at four different resistivity values
    (i.e $\eta = 3 \times 10^{-4},  5 \times 10^{-4},  2 \times 10^{-3}$, and $6 \times 10^{-3}$). The expected Sweet-Parker (SP1/SP2) scalings in $(1 + P_m)^{-1/2}$,
     $(1 + P_m)^{-1/4}$, and  $(1 + P_m)^{-3/4}$, for $V_s$, $J_M$, and $\Omega_M$ respectively, are also plotted for comparison. Additional power laws following $P_m^{-1/4} - P_m^{-0.4}$
      dependences for $V_s$ are also plotted.
     }
   \label{fig13}
\end{figure}

\

\end{document}